\documentclass{article}

\usepackage[utf8]{inputenc}%
\usepackage{xcolor}
\def\blue{\color{blue}}

\usepackage{graphicx}
\usepackage{hyperref}
\usepackage{amsmath}

\usepackage{comment} %

\usepackage{dcolumn} %
\usepackage{paralist} %

\usepackage[font={small,sf}]{caption}

\usepackage[sf]{titlesec}
\titleformat*{\section}{\sf\blue\LARGE}
\titleformat*{\subsection}{\sf\blue\Large}
\titleformat*{\subsubsection}{\sf\blue\large}
\titleformat*{\paragraph}{\sf\blue}
\titleformat*{\subparagraph}{\sf\blue}
\usepackage{soul}

\def\sebastian#1{\textcolor{red}{[[#1 -- sebastian]]}}

\def\chkAtEnd{\relax}
\ifdefined \user
\usepackage{geometry}
\else
\usepackage{geometry}
\fi

\def\calibrationParam{\Theta}
\def\shedding{sh}
\def\contactIntensity{ci}
\def\contactIntensityReduced{ci'}
\def\intake{in}
\def\duration{\tau}
\def\roomSize{rs}
\def\airExchange{ae}
\def\facilityFloorSpace{facilityFloorSpace}
\def\maxPersonsAtFacility{N^{\it personsAtFacility}_{\it max}}
\def\typicalFloorSpacePerPerson{{\it fs}}
\def\nSpacesPerFacility{N^{\it spacesPerFacility}}
\def\effectiveMaxPersonsAtFacility{N^{\it personsInRoom}_{\it max}}

\def\showingSymptoms{\emph{showingSymptoms}}
\def\seriouslySick{\emph{seriouslySick}}
\def\critical{\emph{critical}}

\usepackage{hyperref}
\hypersetup{colorlinks=true}
\begin{document}
\title{\sf\blue A realistic agent-based simulation model for COVID-19 based on a traffic simulation and mobile phone data}
\author{%
Sebastian A.\ Müller${}^{1*}$, Michael Balmer${}^2$, William Charlton${}^1$, Ricardo Ewert${}^1$, 
\\
Andreas Neumann${}^3$, Christian Rakow${}^1$, Tilmann Schlenther${}^1$, Kai Nagel${}^{1*}$
\\
\footnotesize ${}^1$ Transport Systems Planning and Transport Telematics, TU Berlin, Germany
\\
\footnotesize ${}^2$ Senozon AG, Switzerland
\\
\footnotesize ${}^3$ Senozon GmbH, Germany
\\
\footnotesize ${}^*$ corresponding authors
}
\maketitle

\begin{abstract}\noindent
Epidemiological simulations as a method are used to better understand and predict the spreading of infectious diseases, for example of COVID-19.  
This paper presents an approach that combines person-centric data-driven human mobility modelling with a mechanistic infection model and a person-centric disease progression model.  
The model includes the consequences of disease import, of changed activity participation rates over time (coming from mobility data), of masks, of indoors vs.\ outdoors leisure activities, and of contact tracing.

Results show that the model is able to credibly track the infection dynamics in Berlin (Germany).  The model can be used to understand the contributions of different activity types to the infection dynamics over time.  The model clearly shows the effects of contact reductions, school closures/vacations, or the effect of moving leisure activities from outdoors to indoors in fall.  Sensitivity tests show that all ingredients of the model are necessary to track the current infection dynamics.
One interesting result from the mobility data is that behavioral changes of the population mostly happened \textit{before} the government-initiated so-called contact ban came into effect.  Similarly, people started drifting back to their normal activity patterns \emph{before} the government officially reduced the contact ban.
Our work shows that is is possible to build detailed epidemiological simulations from microscopic mobility models relatively quickly.  They can be used to investigate mechanical aspects of the dynamics, such as the transmission from political decisions via human behavior to infections, consequences of different lockdown measures, consequences of wearing masks in certain situations, or contact tracing.
\end{abstract}

\section{Introduction}
\label{sec:introduction}

The general dynamics of virus spreading is captured by compartmental models, most famously the so-called SIR model, with  $S$  = \textit{susceptible},  $I$  = \textit{infected/infectious}, and  $R$  = \textit{recovered} \cite{Kermack1927-go,Anderson1979-oj}.  Every time a \textit{susceptible} and an \textit{infectious} person meet, there is a probability that the susceptible person becomes infected. Some time after the infection, the person typically recovers.  Variants include, e.g., an \textit{exposed} (but not yet \textit{infectious}) compartment between \textit{S} and \textit{I}.

Instead of running these models with compartments, one can run them on a graph \cite{Bajardi2011-pe,Belik2011-ru,Iannelli2017-wa,Leventhal2015-gb,Pastor-Satorras2015-do}.  Persons are represented as vertices, connections between persons are denoted as edges.  The random interactions that are implied by the compartmental models are then replaced by interactions with graph neighbors.

In reality, these interactions change from day to day; in particular, possible superspreading events like weddings or other large gatherings cannot be encoded in a static graph.  For this, temporal networks have been investigated (\cite{Pastor-Satorras2015-do}, section~VIII).

Finally, a ``different framework emerges if we consider nodes as entities where multiple individuals or particles can be located and eventually wander by moving along the links connecting the nodes'' \cite{Pastor-Satorras2015-do}.  When COVID-19 took hold in Europe, one model of this type by Imperial College \cite{Halloran2008-ag,Ferguson2020-lk} had a large impact on policy in the UK.  Other examples of this approach are 
by the Virginia Biotechnology Institute \cite{Eubank2004-ri,Halloran2008-ag} 
and by the Center for Statistics and Quantitative Infectious Diseases in Seattle \cite{Chao2010-hx,Halloran2008-ag}.  
Examples for similar approaches on the global level are \cite{Hufnagel2004-af,Bajardi2011-pe}.
Groups that started more recently include \cite{Davids2020-dh} and \cite{Tadic2020-zi}.  

Our own approach in this direction, presented in this paper, continues work by Smieszek et al \cite{ 
Smieszek2009-vw,Smieszek2011-hc} and by Hackl and Dubernet \cite{Hackl2019-ix}.  
The important difference, and major innovation, is that our model is entirely data-driven on the mobility side, i.e. both the ``normal''  person trajectories and the reduction of activity participation over the course of the epidemics stem from data. This allows to considerably speed up the model implementatation, and to reduce the number of free parameters. 
In the present paper, we show how such a system was built up 
rather quickly from pre-existing synthetic mobility traces from mobile phone data, which were originally generated for traffic applications.  In fact, we built a prototype in about two weeks \cite{MuellerEtAl2020COVID-19}.  Subsequently, we received funding to continue our research and to regularly report to the ministry of research of Germany (e.g.~\cite{Muller2020-vq,Muller2020-ks}). 

The model is used to replay the epidemics in Berlin.  This allows important insights in the transmission from government actions to mobility behavior to infection dynamics.  Importantly, it will turn out that at least in Berlin and presumably in Germany, the population started reducing its out-of-home activities before the government asked/ordered the population to do so. 

\section{Model details}
\label{sec:model}

Important sub-models of agent-based epidemics models are: contact model, infection model, and disease progression model.  These are described in more detail in the following sections.

\subsection{Contact model}
\label{sec:contactModel}

\label{sec:contactModelGeneral}

As stated, we take the contact model from transport modelling, more precisely from activity-based transport modelling (e.g.~\cite{Axhausen1989-wn,BowmanEtc1999PortlandActs,Hilgert2017-di}).  Such models generate complete daily activity chains of persons, for example something like home–work–shop–home–leisure–home. Activities come with times and, importantly, locations. The activity chains are normally used as input for (agent based) transport simulations, which assign modes and routes and possibly re-adjust times and locations, and thus generate emergent effects such as congestion and emissions \cite{Horni2016-ge}. 
In the present paper, they are instead used as input to an epidemic model.

\begin{figure}[!h!t!]
\centering
\includegraphics[width=0.8\textwidth,trim=3.5cm 6cm 2.5cm 5cm,clip]{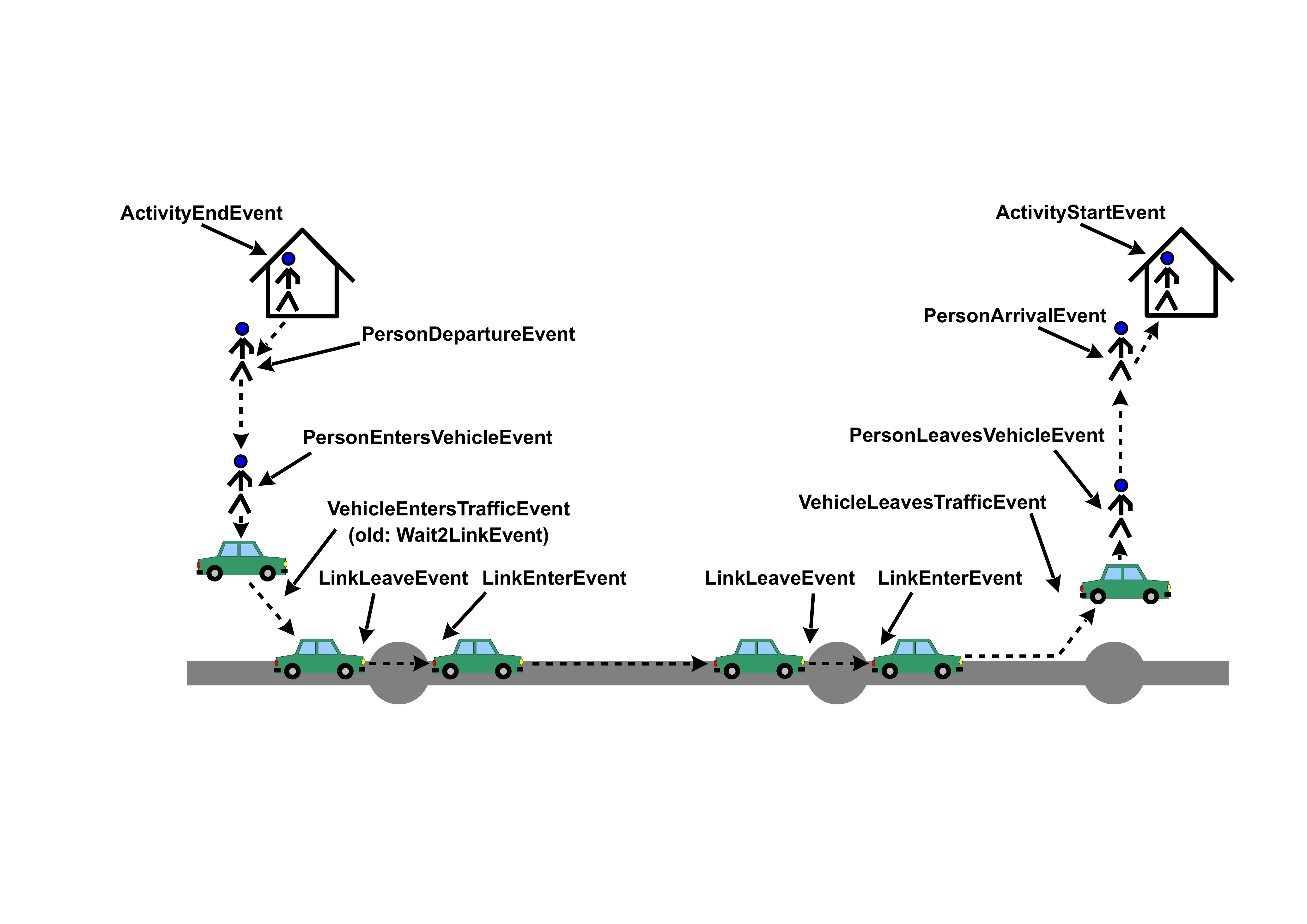}
\includegraphics[width=0.8\textwidth,trim=3.5cm 6cm 2.5cm 5cm,clip]{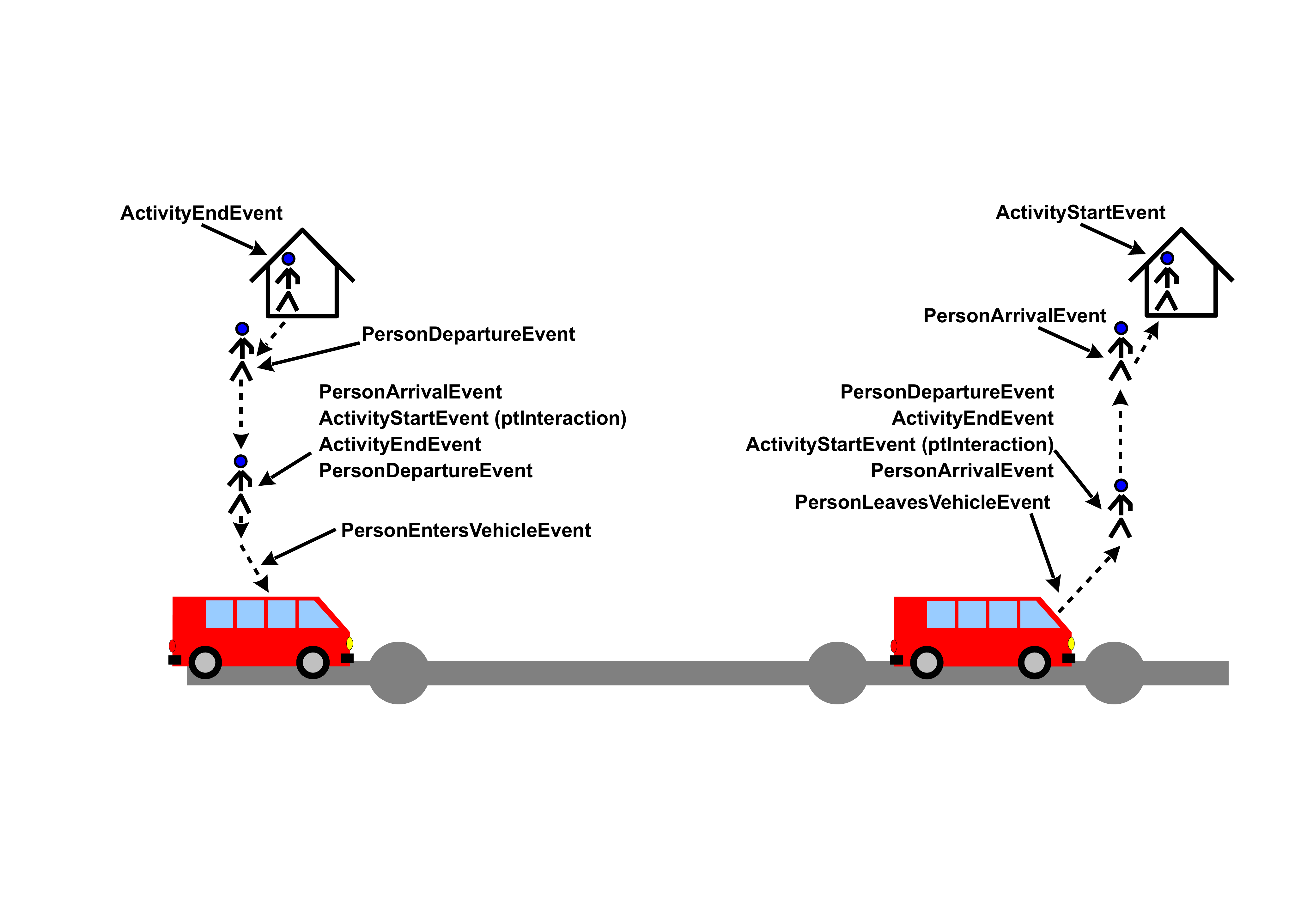}
\caption{TOP: Events for travel by individual vehicle.  BOTTOM: Events for travel by public transport.  Source: \cite{ Horni2016-ge}.}
\label{fig:events}
\end{figure}

For these activity chains, one could, for example, use trajectories from mobile phone data \cite{PaniguttiTizzoniBajardiEtAl2017MobilePhoneDataForMobilityPatternsInSpatialEpidemics}.  These trajectories are often not available, for example for privacy reasons. It is, however, possible to generate synthetic approximations to these trajectories.  One approach is to use information from mobile phone data (but not the full trajectories), and process them together with information about the transport system and with statistical information from other surveys \cite{ Senozon2020-xh,Senozon2020-or}.  That approach leads to synthetic movement trajectories for the complete population (cf.~Fig.~\ref{fig:events}).  

From these trajectories, we extract how much time people spend with other people at activities or in (public transport) vehicles.  That is, infection opportunities are directly taken from the input data.

\paragraph{Handling of large facilities}
\label{sec:subSpaces}

The resolution of our input data comes at the level of ``facilities''.  Those can be interpreted as buildings or sometimes blocks.  They often contain multiple households, multiple company offices, multiple leisure facilities, multiple shops, etc.  
For \emph{home} activities, we split persons living in the same facility into realistic household sizes with a maximum number of six people per household \cite{Destatis2020-og}.  This seems important since the within-household dynamics of COVID-19, and in particular the fact that the secondary attack rate in households seems to be far below 100\%, plays an important role (e.g.~\cite{Streeck2020-sm}).
For all other activities, we divide the facilities by some globally set factor, called $\nSpacesPerFacility$.  That is, if two persons spend overlapping time at the same facility, the probability that they have interacted is $1/\nSpacesPerFacility$.  This has important ramifications for multi-day modelling and mixing, see 
below.

\paragraph{Multi-day modelling}
\label{sec:multi-day-modelling}

Optimally, one would have multi-day trajectories.  In our case, the data that we have ends at the end of the day.  Our simulations thus run the same person trajectories again and again (except for weekends, see %
below).  This presumably \textit{underestimates} mixing, since it is plausible to assume that there is some variation in activity patterns from day to day.  
At this point, one needs to make a decision whether our sub-spaces %
(see above) are frozen, meaning that the same sub-groups meet every day, or not.  Using the same sub-groups every day arguably is plausible for office buildings, which may contain offices for several companies, and interaction may be limited to sharing an elevator.  It is less plausible for public transport trains, where passengers are arranged differently every day.  Possibly, a mix between the two approaches is plausible, introducing the need for even more free parameters.  In our present model, we opt for the non-frozen setting, i.e.\ the other persons within a facility that an ego person interacts with are randomly re-drawn for every new simulated day.

$\nSpacesPerFacility$ evidently influences the number of contacts that a person has.  For our simulations, we set it such that that number of contacts is roughly consistent with real-world contact tracing.  For our current input data, that leads to a setting of $\nSpacesPerFacility=20$\chkAtEnd.

\paragraph{Weekend modelling}
\label{sec:weekend}

As already alluded to 
above, we use separate models for Saturdays and Sundays.  They come out of transport modelling in the same way as we obtain the model for a ``typical weekday'' (see above).
These models use the same synthetic persons and facilities, and thus can be aligned with the weekday model.  In consequence, each synthetic person in our models, starting on monday,
\begin{inparaenum}[(a)]
\item repeats the same weekday five times,
\item runs her Saturday schedule,
\item runs her Sunday schedule,
\end{inparaenum}
and then starts over.

\paragraph{25\% sample}

For computational reasons, we use a 25\% sample of the full population.  The sample is constructed by choosing 25\% of all persons in the population randomly and retaining their full trajectories. The splitting of households as described 
above is done \emph{after} the sampling, meaning that we have realistic household sizes in the 25\% scenario but consider only 25\% of them; also, the number of contacts to determine the parameter $\nSpacesPerFacility$ %
(see above) 
is determined for the 25\% model. 
We have also run the full 100\% model to check that there are no major differences.  The 25\% model allows to finish runs within a single-digit number of hours, which was and is important for fast model turn-around driven by the the necessity for fast progress given the demand for the results by the decisionmakers.
All results are reported after upscaling to 100\%.

\subsection{Infection model}

\label{sec:infection model}
Once two persons are identified to have contact, and one of them is contagious and the other is susceptible, there is a probability of an infection.  For this, we use a mechanical model by Smieszek \cite{Smieszek2009-vw,Smieszek2010ModelsOfEpidemicsPhD}: Infected persons generate a ``viral load''  that they exhale, cough or sneeze into the environment, and people close by are exposed.  Overall, the probability for person $n$ to become infected by this process in a time step  $t$  is described as
\begin{equation}\boxed{%
p(infect|contact)=1-\exp  \left( - \calibrationParam  \sum _{m}^{}\shedding_{m,t_{}}\cdot  \contactIntensity_{nm,t_{}}\cdot \intake_{n,t}~\cdot  \duration_{nm,t_{}}\right)
\label{eq:infect}
}\end{equation}
where  $m$  is a sum over all other persons, $\shedding$ is the shedding rate ($\sim$ microbial load),  $\contactIntensity$  the contact intensity,  $\intake$  the intake (reduced, e.g., by a mask),  $\duration$  the duration of interaction between the two individuals, and $\calibrationParam$ a calibration parameter.\footnote{%
Note that this is structurally the same as a continuous time dynamics with rate $\beta$ and a limited infective period: The probability to not become infected during $\delta t$ is $1-\beta \, \delta t$; if there are $\tau/\delta t$ subsequent such periods, that probability becomes $(1-\beta \, \delta t)^{\tau/\delta t}$, and the probability to \emph{become} infected thus becomes \cite{Pastor-Satorras2015-do} (section V.4)
$
1 - \lim_{\delta t \to 0} (1 - \beta \, \delta t)^{\tau/\delta t} = 1 - e^{-\beta \, \tau} \ .
$
However, in our case $\tau$ only goes from the beginning to the end of the overlap, and is repeated every day again, given that both persons are still in the same states.
}

For small values of the exponent, one can approximate Eq.~(\ref{eq:infect}) as
\begin{equation}\boxed{%
p(infect|contact) \approx \calibrationParam \times \shedding \times \contactIntensity \times \intake \times \duration \ .
\label{eq:infect-lin}  
}\end{equation}
We do not use this approximation in our computer implementation, but it helps understanding the following arguments.

All parameters can be given in arbitrary units as long as they are always the same since the units are absorbed by $\calibrationParam$.

\paragraph{Contact intensities}

For SARS-CoV-2, it is plausible to assume that a large share of the virus material is shed as aerosol \cite{Marr_et_al2020-ns}.  In consequence, the first relevant term to compute the viral concentration in the air is the shedding rate, $\shedding$.

For such aerosols, it is plausible to assume that they mix quickly into the room, leading to the same uniform concentration everywhere \cite{Hartmann2020-vx}.  Evidently, that concentration is indirectly proportional to room size: if the room is twice as large, the resulting concentration is half as large.

Next, air exchange plays a role.  One could, for example, assume that the windows are opened once per hour, and all of the air is replaced with outside air.  This would correspond to an air exchange rate of 1/h.  If one assumes a constant rate of virus emission, there would be a linear increase of concentration up to the opening of the window, after which (in a theoretical model) the virus concentration in the air would quickly drop to zero.  The \emph{average} virus concentration over this process would be half as much as the maximum concentration just before window opening.  In consequence, the resulting average concentration is indirectly proportional to the air exchange rate: If the air is exchanged twice as often, the resulting average virus concentration is half as large.  This also holds for continuous air exchange, e.g.\ by mechanical means.

All of the above together replaces Eq.~\ref{eq:infect-lin} by
\begin{equation}\boxed{%
p(infect|contact) \approx \calibrationParam \, \frac{\shedding \times \intake}{\roomSize \times \airExchange} \, \duration  \ ,
}\end{equation}
where $\roomSize$ is the size of the room, and $\airExchange$ is the air exchange rate.  That is, it sets
\begin{equation}
\contactIntensity = \frac{1}{\roomSize \times \airExchange} \ .
\label{eq:ciRoomsizeAirexchange}
\end{equation}
Again, the physical units are absorbed into $\Theta$; note, however, that the air exchange rate $\airExchange$ is defined as exchanging air for the full room, and not in, say, cubic meters.  

\paragraph{Estimation of room sizes}

As stated above, our data resolves down to the level of ``facilities''.  These correspond roughly to buildings.  In consequence, such a facility can be anything from a single family home to a large office building to a sports arena.  

Since our simulation tracks when persons are at facilities, we can, for each facility, obtain the maximum number of persons at that facility, $\maxPersonsAtFacility$, over the day.  In addition, one can obtain typical floor space per person, $\typicalFloorSpacePerPerson$, from regulatory norms and other sources (see Sec.~\ref{sec:ci-concreteNumbers}).  This leads to
\begin{equation}
\facilityFloorSpace = \maxPersonsAtFacility \times \typicalFloorSpacePerPerson \ .
\label{eq:floorSpace}
\end{equation}
Since we divide all facilities by $\nSpacesPerFacility$, this leads for the room size to
\begin{equation}
\roomSize = 
\effectiveMaxPersonsAtFacility \times \typicalFloorSpacePerPerson 
\label{eq:roomsize}
\end{equation}
with
\begin{equation}
\effectiveMaxPersonsAtFacility := \frac{\maxPersonsAtFacility}{\nSpacesPerFacility} ;  
\label{eq:maxPersonsInRoom}
\end{equation}
recall that $\nSpacesPerFacility=1$ for home activities.

\paragraph{Concrete numbers}
\label{sec:ci-concreteNumbers}

Overall, the above results in
\begin{equation}
\contactIntensity = 
\frac{1}{\effectiveMaxPersonsAtFacility \times \typicalFloorSpacePerPerson \times \airExchange} \ ,
\end{equation}
which can be re-written as
\begin{equation}
\contactIntensity = \frac{1}{\effectiveMaxPersonsAtFacility} \times \contactIntensityReduced
\label{eq:ciRewritten}
\end{equation}
with the ``normalized'' contact intensity
\begin{equation}
\contactIntensityReduced = \frac{1}{\typicalFloorSpacePerPerson \times \airExchange} \ .
\label{eq:ciPrime}
\end{equation}
See Table~\ref{tab:contactIntensities} for values of $ci'$.

Although Eqs.~(\ref{eq:ciRewritten}) and (\ref{eq:ciPrime}) really mean the same thing as Eqs.~(\ref{eq:ciRoomsizeAirexchange}) to (\ref{eq:maxPersonsInRoom}), we find them more difficult to interpret.  $\contactIntensityReduced$ is the easy part -- it parameterizes the ``closeness'' of the interaction.  But why should that number be divided by $\effectiveMaxPersonsAtFacility$, as Eq.~(\ref{eq:ciRewritten}) implies?  The reason is that facilities/rooms that are used simultaneously by a larger number of people are also expected to be larger.  And, from a personal perspective, if we share a room with one infectious other person, then our probability to become infected is, all other things being equal, indeed half as large if the room is twice as large.  
However, when it is twice as large, then there will presumably also be twice as many persons in it, doubling our own risk, and thus in the average cancelling out the effect of the larger room size.  This second effect, however, is computed directly by our contact model (Sec.~\ref{sec:contactModelGeneral}), and thus does not have to be included into the contact intensity.  This has the additional advantage that if a person is in large container outside its peak usage, the model will calculate a much reduced infection probability.  Examples for this are public transport vehicles, premises for large events, or restaurants.

\begin{table}[!h!]
\centering
	\caption{Normalized contact intensities $\contactIntensityReduced$. 
The floor area per person and the air exchange rate come from building manuals or similar standards. 
The share of old buildings/vehicles is an estimate.  
Universities are assumed to have twice as much space per student as schools.
Shop, errands, and business are assumed to follow the same characteristics as work.
The resulting normalized contact intensity is computed from the other values as
$
\contactIntensity = \langle 1/(\typicalFloorSpacePerPerson \times \airExchange) \rangle \ ,
$
where $\langle . \rangle$ denotes the average over old and new buildings/vehicles that we use for Berlin.
}
	\label{tab:contactIntensities}
	\vspace{-1em}
	\def\mc#1{\multicolumn{1}{c}{#1}}
		\begin{tabular}{r D{.}{.}{3.2} D{.}{.}{3.2} D{.}{.}{3.2} r D{.}{.}{3.2}}
			& & & & & \mc{} \\
			\mc{activity} & \mc{area per} & \mc{air exchange} & \mc{air exchange} & \mc{share old} & \mc{resulting} \\
			\mc{type} & \mc{person} & \mc{rate low} & \mc{rate high} & \mc{buildings} & \mc{$\contactIntensityReduced$} \\
& \mc{[$m^2$]} & \mc{[$1/h$]} & \mc{[$1/h$]} & \mc{/vehicles}& \\ \hline
			home \cite{Senatsverwaltung_fur_Integration2020-fp} & 22 & 0.5 & 0.5 & & 1 \\
			schools and day care \cite{Wolters_Kluwer_Deutschland_GmbH2020-wc} & 2 & 0.5 & 0.5 & 100\% & 11 \\
			universities & 4 & 0.5 & 0.5 & 100\% & 5.5 \\
			public transport & 0.33 & 2.0 & 10.0 & 50\% & 10 \\
			leisure \cite{Des_Ausschusses_fur_staatlichen_Hochbau_der_Bauministerkonferenz2002-pe} & 1.25 & 0.5 & 10.0 & 50\% & 9.24 \\
			shop & 10 & 0.5 & 1.5 & 10\% & 0.88 \\
			work \cite{2020-uy,noauthor_2012-yq,DIN_Deutsches_Institut_fur_Normung2015-se} & 10 & 0.5 & 1.5 & 50\% & 1.47 \\
			errands & 10 & 0.5 & 1.5 & 50\% & 1.47 \\
			business & 10 & 0.5 & 1.5 & 50\% & 1.47 \\\hline
		\end{tabular}
\end{table}

\paragraph{Children}
\label{sec:children}

Current research implies that the susceptibility and infectivity are reduced for children compared to adults. We model this by including the susceptibility and infectivity into Eq.~(\ref{eq:infect}). For adults both parameters are set to one. For people below the age of twenty the infectivity is reduced to 0.85 and the susceptibility to 0.45 \cite{Dattner2020-ug}. Note that this does not mean that the infection probability for children is necessarily lower than for adults because children are more likely to perform activities with a high contact intensity (as shown in Table~\ref{tab:contactIntensities}; also see Sec.~\ref{sec:reduced-infectivity-etc} in the appendix).  

\subsection{Disease progression model}
\label{sec:dise-progr-model}

\begin{figure}[!t!]
\centering
\includegraphics[width=0.8\hsize,trim=5cm 0 0 0,clip]{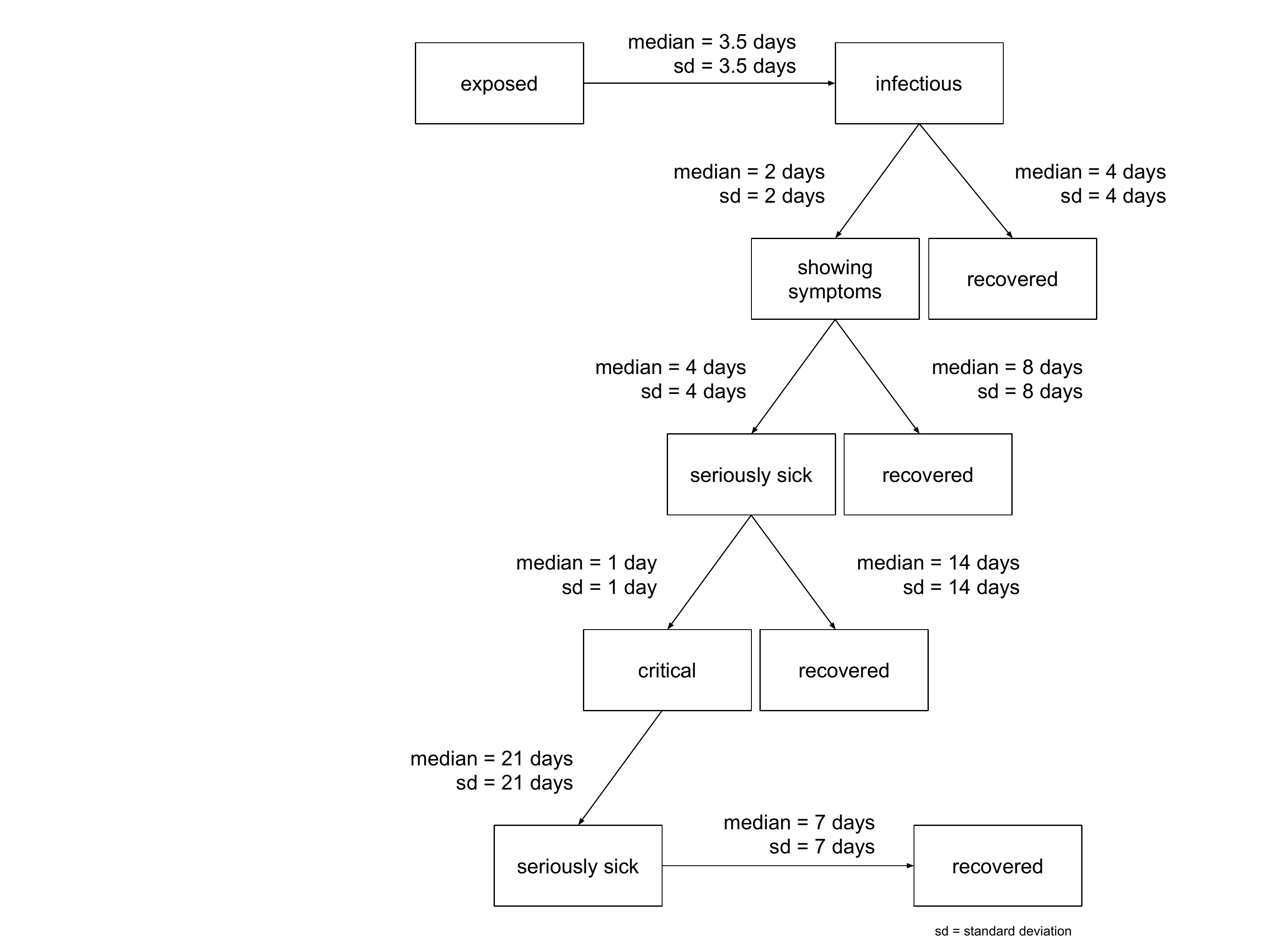}
\caption{State transitions  \cite{Who2020-ys,He2020-mm,Wolfel2020-vr,Dreher2020-rq,Wang2020-bd,Robert_Koch_Institute2020-sg}. 
}
\label{fig:stateTransitions}
\end{figure}

The disease progression model is taken from the literature \cite{Who2020-ys,He2020-mm,Wolfel2020-vr,Dreher2020-rq,Wang2020-bd,Robert_Koch_Institute2020-sg} (also see \cite{Ashcroft2020-wo}).  
The model has states \textit{exposed, infectious, showing symptoms, seriously sick}($=$ should be in hospital), \textit{critical} ($=$ needs intensive care), and \textit{recovered}. 
The durations from one state to the next follow log-normal distributions; 
see Fig.~\ref{fig:stateTransitions} for details.
We use the same age-dependent transition probabilities as \cite{Ferguson2020-lk}, shown in Tab.~\ref{tab:ageDepTransProbas}.

Infecting another person is possible during \textit{infectious}, and while \textit{showing symptoms}, but no longer than 4 days after becoming infectious.  This models that persons are mostly infectious relatively early through the disease \cite{ He2020-mm}, while in later stages the infection may move to the lung \cite{ Wolfel2020-vr}, which makes it worse for the infected person, but seems to make it less infectious to other persons.  

\begin{table}[!h!t!]
\centering
\caption{Age-dependent transition probabilities from symptomatic to seriously sick (= requiring hospitalisation), and from seriously sick to critical (= requiring breathing support or intensive care).  Source: \cite{ Ferguson2020-lk}.}
\label{tab:ageDepTransProbas}
\begin{tabular}{ccc}
\hline
Age-group & symptomatic cases & hospitalised cases \\
& requiring hospitalisation & requiring critical care \\
\hline
0 to 9 & 0.1\% & 5.0\% \\
10 to 19 & 0.3\% & 5.0\% \\
20 to 29 & 1.2\% & 5.0\% \\
30 to 39 & 3.2\% & 5.0\% \\
40 to 49 & 4.9\% & 6.3\% \\
50 to 59 & 10.2\% & 12.2\% \\
60 to 69 & 16.6\% & 27.4\% \\
70 to 79 & 24.3\% & 43.2\% \\
80+ & 27.3\% & 70.9\% \\
\hline
\end{tabular}
\end{table}

\subsection{Simulation runs}

This paper presents simulation results for the metropolitan area of Berlin in Germany, with approx.~5~million people.  A typical simulation run looks as follows:
\begin{enumerate}
	\item One or more \textit{exposed} persons are introduced into the population.

	\item At some point, \textit{exposed} persons become \textit{infectious}.  From then on, every time they spend time together with some other person in a vehicle or at some activity, Eq.~(\ref{eq:infect}) is used to calculate the probability that the other person, if \textit{susceptible}, can become infected (= \textit{exposed}).  If infection happens, the newly infected person will follow the same progression.

\item \textit{Infectious} persons eventually move on to other states, as described in Fig.~\ref{fig:stateTransitions}.
\end{enumerate}
The model runs many days, until no more infections occur.  

\section{Methods and Results}

\subsection{Base calibration}
\label{sec:base-calibration}

\begin{figure}[!h!]
\centering
\includegraphics[width=0.8\textwidth]{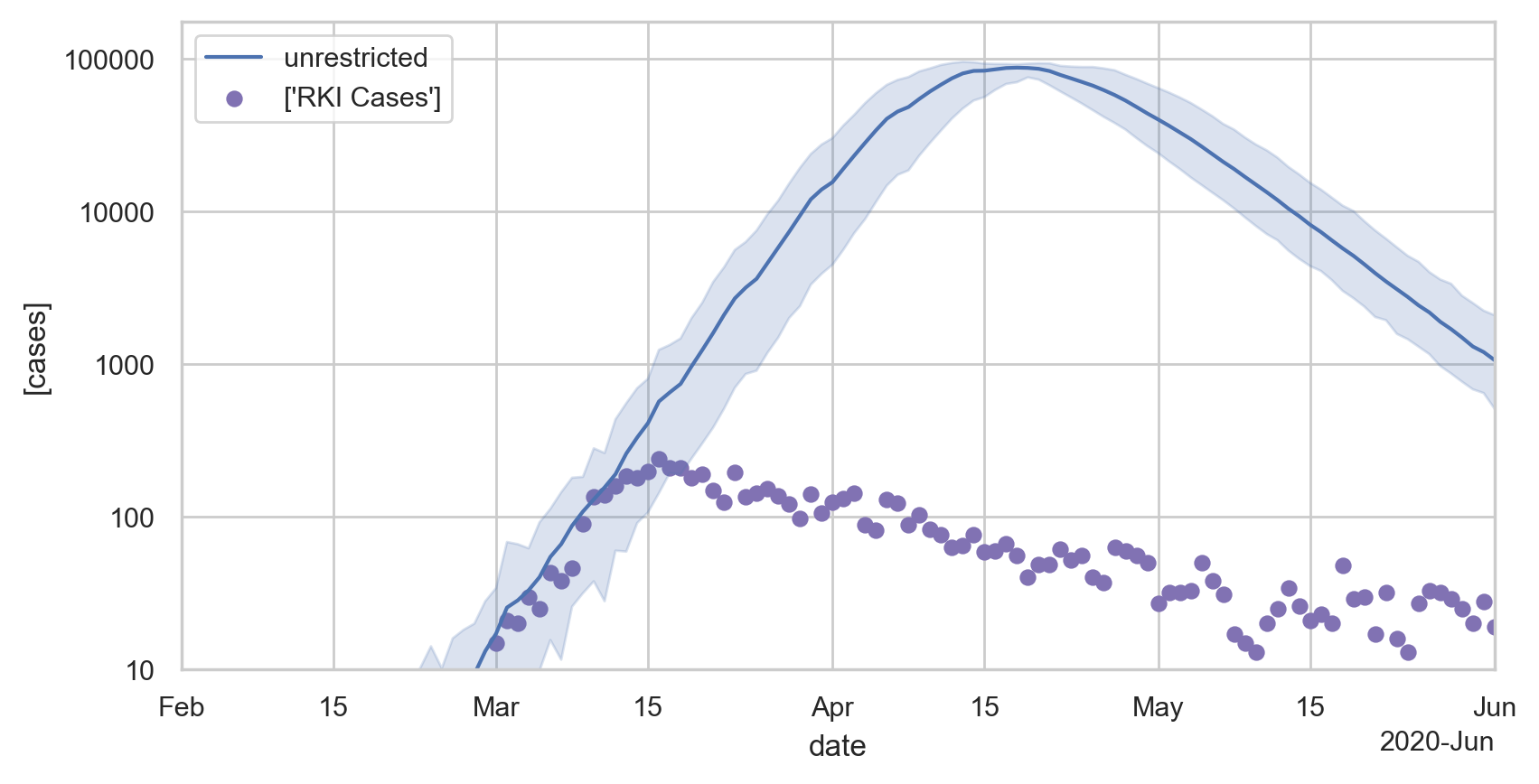}
\caption{Unrestricted base case.  The blue line is averaged over 10~independent Monte Carlo runs with different random seeds.  The blue dots denote case numbers as reported by Robert Koch Institute~\cite{Robert_Koch_Institute2020-jp}.}
\label{fig:infections-unrestricted}
\end{figure}

Most parameters of the model are taken from the literature, as explained in Sec.~\ref{sec:dise-progr-model}.  The remaining free parameters are, from Eq.~(\ref{eq:infect}),  $\calibrationParam$, $\shedding$, and $\intake$.  We have 
	 set the base values of $\shedding=\intake=1$.  As mentioned in Section~\ref{sec:infection model}, we use these parameters to model the wearing of masks, meaning that they are reduced when masks are worn.
The remaining $\calibrationParam$ parameter was then calibrated so that in the base case, the number of cases doubles every three days. This is a plausible fit to the initial days which were totally without restrictions, which is what our base case represents. Calibrating  $\calibrationParam$ in our model corresponds to calibrating $\beta$, the infection rate, in a compartmental model. 
The result can be seen in Fig.~\ref{fig:infections-unrestricted}.

\subsection{Dynamic input data}
\label{sec:dynamicInputData}

As additional elements, we feed into the simulation the elements
\begin{inparaenum}[(a)]
\item disease import,
\item reductions in activity participation, 
\item mask compliance during shopping and in public transport, and
\item the effects of outdoors vs.\ indoors season.
\end{inparaenum}
All of these elements are obtained from data, so we do not need to guess them.

In addition, we add contact tracing followed by quarantine-at-home, since that is important for the behavior of the epidemics in Berlin in fall.

Unfortunately, it is not possible to add these aspects one by one, since every time one aspect is added, the value of $\calibrationParam$ needs to be re-calibrated.  We will therefore explain all four aspects in the following subsections, and then show model runs where all four elements are used for the model.  Sensitivity tests, where each of these elements is individually removed, are provided in the appendix (Sec.~\ref{sec:sensitivities}).

\subsubsection{Disease import}
\label{sec:diseaseImport}

\begin{figure}
    \centering
    \includegraphics[width=0.8\textwidth]{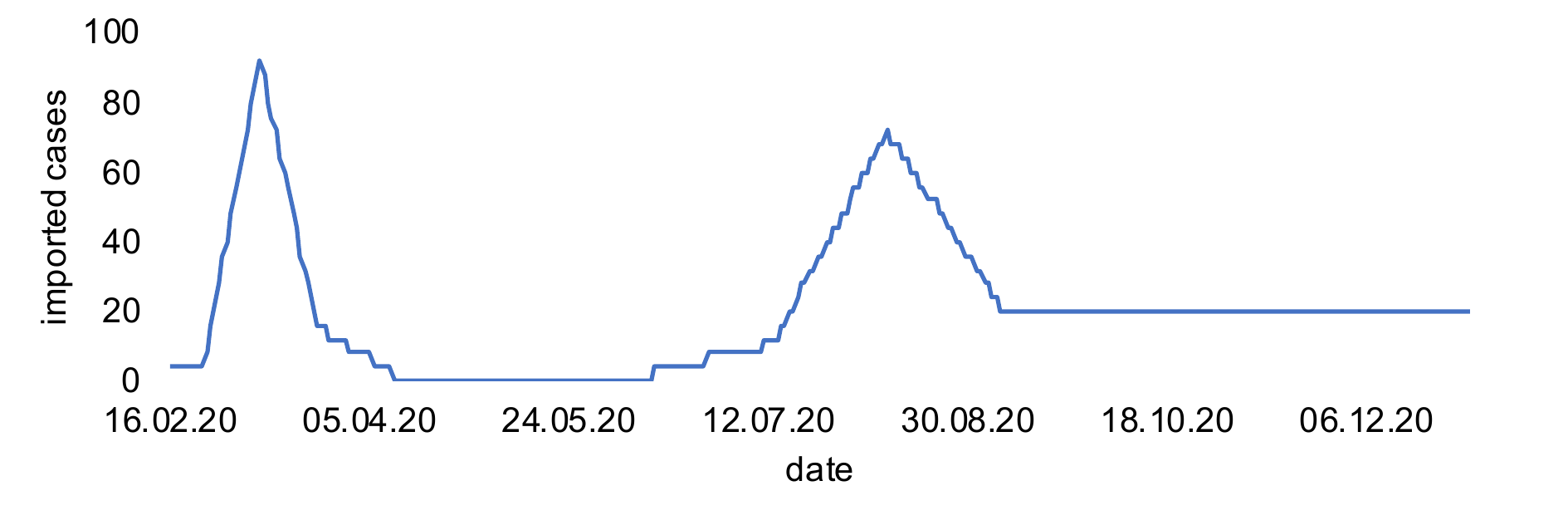}
    \caption{Disease import over time.  Data taken from \cite{Robert_Koch_Institute2020-jp} (always on tuesdays).}
    \label{fig:disease-import}
\end{figure}

We take the disease import from abroad from data published by RKI (\cite{Robert_Koch_Institute2020-jp}, always on tuesdays). Currently, for Germany this data is only available on a nationwide aggregated level. For this reason we scale it down to our Berlin model by using the population size. The data is dated on the reporting date and not on the actual date of becoming sick. Since the infection seeds are initiated into our model with the status exposed (see Fig.~\ref{fig:stateTransitions}) and it can be assumed that the reporting date is significantly after the exposure date we date the data from RKI back by one week.
The data provided by RKI is available as weekly values so we assign these values to the respective monday and then interpolate between them. The initially infected persons are drawn randomly from the population. The resulting disease import is shown in Fig.~\ref{fig:disease-import}.

\subsubsection{Reductions in activity participation}
\label{sec:reduct-activ-part}

\paragraph{Approach}

During the unfolding of the epidemics, people decided or were ordered to no longer participate in certain activities.  We model this by removing an activity from a person’s schedule, plus the travel to and from the activity.  In consequence, that person no longer interacts with people at that activity location, and in consequence neither can infect other persons nor can become infected during that activity.  Overall, this reduces contact options, and thus reduces epidemic spread.

\begin{figure}[!h!t!]
\centerline{%
\includegraphics[width=0.99\hsize,trim=0 0 0 0,clip]{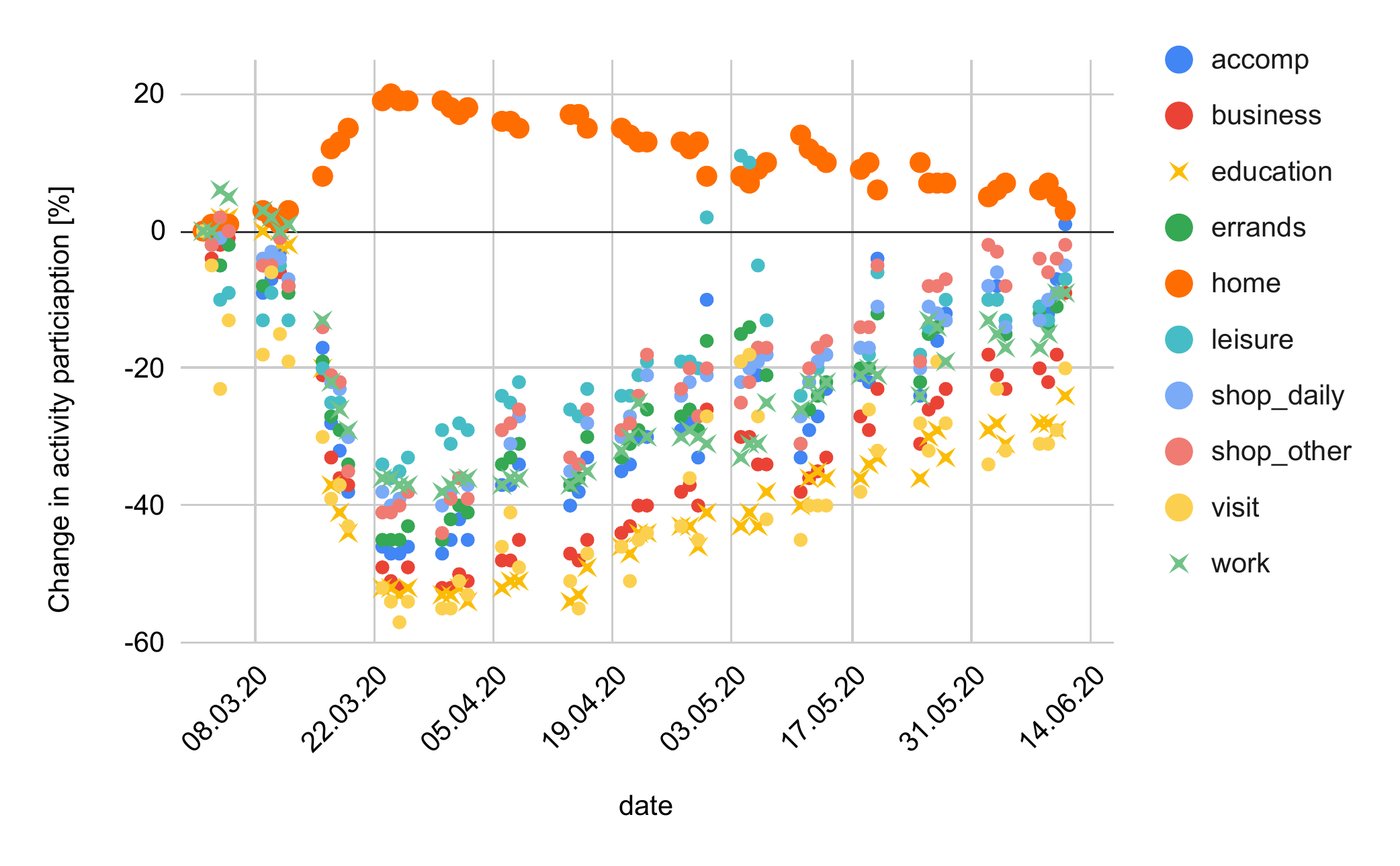}
\hfill
}
\caption{Reduced activity participation over the course of the epidemics in Berlin.}
\label{fig:reducedParticipation1}
\end{figure}

\begin{figure}[!h!t!]
\centerline{%
\includegraphics[width=0.85\hsize,trim=0 0 0 0,clip]{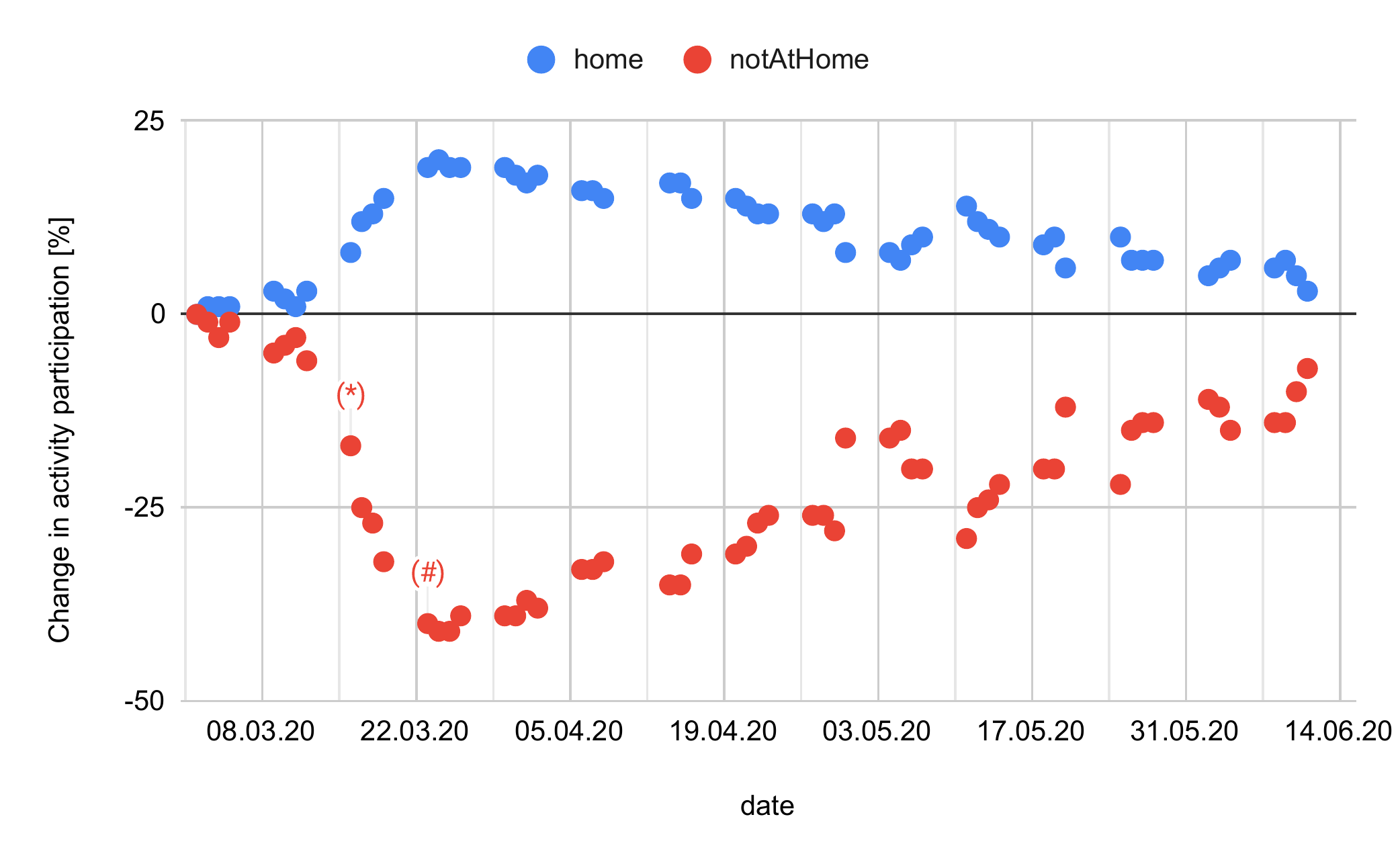}
\hfill}
\caption{Change in activity participation compared to the baseline for normal workdays. All out-of-home activities are combined into one number. (*) denotes the first day of closures of schools, clubs, and bars; and (\#) the first day of the so-called contact ban which came together with closures of all restaurants and non-essential stores.  The data clearly shows that people had reduced their out-of-home activities before the government-ordered closures.}
\label{fig:reducedParticipation2}
\end{figure}

A very important consequence of our modelling approach is that we can take that reduction in activity participation from data.  Unfortunately, the activity type detection algorithm is not very good for these unusual activity patterns, as one can see in Fig.~\ref{fig:reducedParticipation1} when knowing that all educational institutions were closed in Berlin after Mar/15.  What is reliable, though, is the differentiation between at-home and out-of-home time, as displayed in Fig.~\ref{fig:reducedParticipation2}.  One clearly notices that out-of-home activities are somewhat reduced after Mar/8, and dramatically reduced soon after.  After some experimentation, it was decided to take weekly averages of the activity non-participation, and use that uniformly across all activity types in our model, except for educational activities, which were taken as ordered by the government.

To remove an activity with a certain probability, a random draw is made every time a synthetic person has that activity type in its plan.  This means that the model assumes that, say for a 50\% work reduction, there will be another 50\% subset of persons at work every day.  This intervention, in consequence, does not sever infection networks, but just slows down the dynamics.

\paragraph{Behavioral interpretation of government interventions vs mobility data}
\label{sec:behav-interpr-govern}

One striking consequence of the activity participation data (Fig.~\ref{fig:reducedParticipation2}) is that, after the initial government intervention from Mar/7 that cancelled large events and raised awareness, the population reaction in fact \textit{preceded} the government interventions, rather than the other way around. Out-of-home activity participation was already reduced between Mar/7 and Mar/14, \textit{before} the second government intervention that closed schools, clubs, and bars.  Similarly, there was a further considerable drop between Mar/14 and Mar/21, again \textit{before} the so-called contact ban (strongly reduced interaction between different households) and closing of all restaurants and non-essential stores in Germany.  At least for Berlin and probably for Germany, it is thus not true that the government forced society and the economy to come to a halt; rather, society did this by itself, and the government presumably stabilized or reinforced behavior that was happening anyways.  

\subsubsection{Masks}
\label{sec:masks}

In April the wearing of masks in shops and in public transport vehicles became obligatory in Berlin \cite{Wikipedia_contributors_undated-gd}. We have included this into the infection model of Eq.~\ref{eq:infect} by reducing $\shedding$ (if the contagious person wears a mask) and $\intake$ (if the person to be potentially infected wears a mask). This is dependent on the activity type, meaning that persons only wear masks when shopping or using public transport. The effectiveness of different mask types is taken from from \cite{Eikenberry2020-kq}, i.e.\ cloth masks reduce shedding and intake to 0.6 and 0.5 of their original values, surgical masks to 0.3 and 0.3, and N95 (FFP2) masks to 0.15 and 0.025.  
The review article \cite{Chu2020-od} comes up with about 0.05 for N95 masks, a factor of two larger, but still displaying a very large reduction.  The same paper \cite{Chu2020-od} also shows that ``masks'' without a specification of the type has much less of an effect.  Finally, there may be the issue that lay people may not be able to use N95 masks at full efficiency.  In consequence, any of our results that depend on N95 mask efficiency have to be interpreted 
``mechanically'': They are plausible under the assumption that the fraction of people specified in the model is indeed able to use N95 masks effectively.

The local transport company in Berlin (BVG, \cite{Bvg2020-dk}) have provided us with the compliance rates in public transport over time meaning that we do not have to estimate them. We assume that the same compliance rates also apply to shopping activities. We assume that 90\% of those people wearing masks wear cloth masks and 10\% wear N95 masks. 

\begin{figure}
    \centering
    \includegraphics[width=0.8\textwidth]{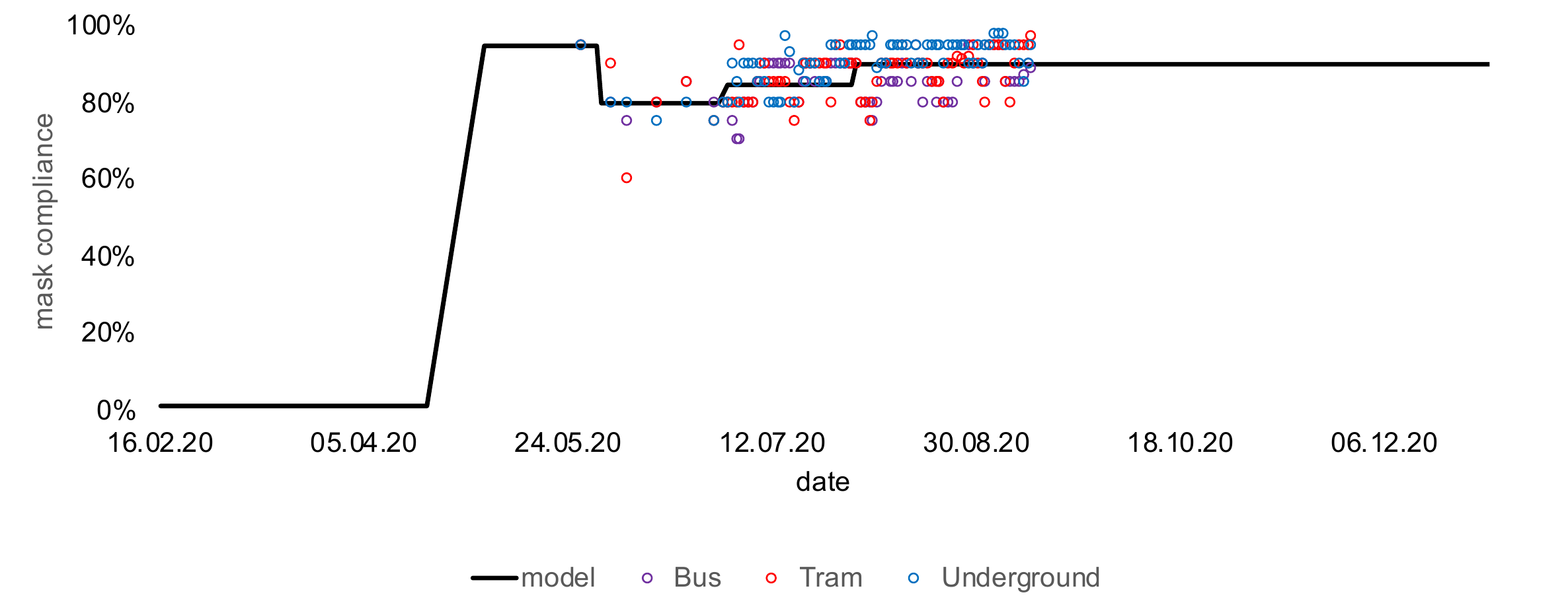}
    \caption{Mask compliance rates over time.}
    \label{fig:mask-compliance}
\end{figure}

\subsubsection{Outdoors vs.\ indoors season}
\label{sec:outdoors-vs-indoors}

The probability of getting infected during an encounter depends on whether the encounter takes place indoors or outdoors. Outside, the probability of infection is significantly reduced compared to inside. This is due to the fact that outdoors the air is constantly in motion and therefore aerosols cannot accumulate. We assume that an encounter outdoors decreases the infection probability by one magnitude \cite{Nishiura2020-es,Marr_et_al2020-ns}.
In countries like Germany, seasonality has a great influence on how much time people spend outside. Based on the German time budget survey \cite{Statistisches_Bundesamt2015-nw} and a survey on physical activities \cite{Statistisches_Bundesamt2017-il} we assume that, in summer, people in Germany spend about 80\% of their leisure time outdoors while this proportion shrinks to 10\% in winter. 
Based on this, 80\% of the leisure activities in our model take place outside from April 15 to September 15. From November 16 until February 15 10\% of leisure activities take place outside.  From September 16 to November 15 and from February 16 to April 15 we interpolate linearly between 80\% and 10\% and 10\% and 80\% respectively.

\subsubsection{Contact tracing}
\label{sec:contactTracing}

The goal of contact tracing is to break chains of transmission by tracing the contacts of an infected person and putting these contacts into quarantine. 
In our model contacts are traced during all activities except for public transport and shopping because we assume that the health authorities are not able to find these contacts. A contact person is only traced when the contact duration is longer than 15 minutes, which corresponds to the RKI guidelines \cite{Robert_Koch_Institut2020-cc}.

Persons that go into \showingSymptoms\ are assumed to trigger a contact tracing mechanism, which works as follows:
\begin{compactenum}
    \item Look at all traced contacts that the infected person had in the 2 days \cite{Robert_Koch_Institut2020-cc} before showing symptoms.
	\item A probability $\gamma$ determines if a contact person can be reached successfully and also follows the stay-at-home order.  $\gamma$ is set to 0.6
	\item The persons that have been traced successfully go into quarantine, but only after a delay of $d$ days, which allows to model the response time of the system. Our base value of $d$ is set to 4 days.  Personal experience in our surroundings says that tests are normally taken a day after symptoms start, and the result is available again one day later in the evening.  That is, contact tracing can start no earlier than 3 days after symptoms onset.  We add another day to account for possible additional delays.
	\item A tracing capacity limits the number of persons per day for which its contacts can be traced. The capacity is set to 
	0 until the end of March,
	30 cases per day until 14/Jun,
	and to 200 cases per day afterwards.  Germany had agreed on a limit of 50 cases per 100\,000 inhabitants per week at which local governments were expected to act \cite{Tagesschau2020-va}.  This number was based on what the system presumably could handle for contact tracing.  For our Berlin scenario with 5~million persons, this translates to 357~cases per day.  Based on newspaper reports \cite{Berliner_Zeitung2020-nv}, the system was overwhelmed already at lower numbers, which is why we use 200.
	\item Persons leave the home quarantine after 14 days, if they did not develop symptoms during that time.
\end{compactenum}
For $d$, a smaller value would be much better in terms of effectiveness, but our personal experience in several cases says that this is unrealistic.  For $\gamma$ and the maximum tracing capacity, we compared simulation results.\footnote{%
\url{https://covid-sim.info/2020-11-09/tracing} 
} Changes in $\gamma$ make relatively little difference.  For the maximum tracing capacity, one can see that larger capacities would have kept the new infections under control for longer than what happened in reality.  

\subsection{Calibration against case numbers in Berlin}
\label{sec:calibr-against-case}

\begin{figure}[!h!t!]
\centerline{\includegraphics[width=0.8\textwidth]{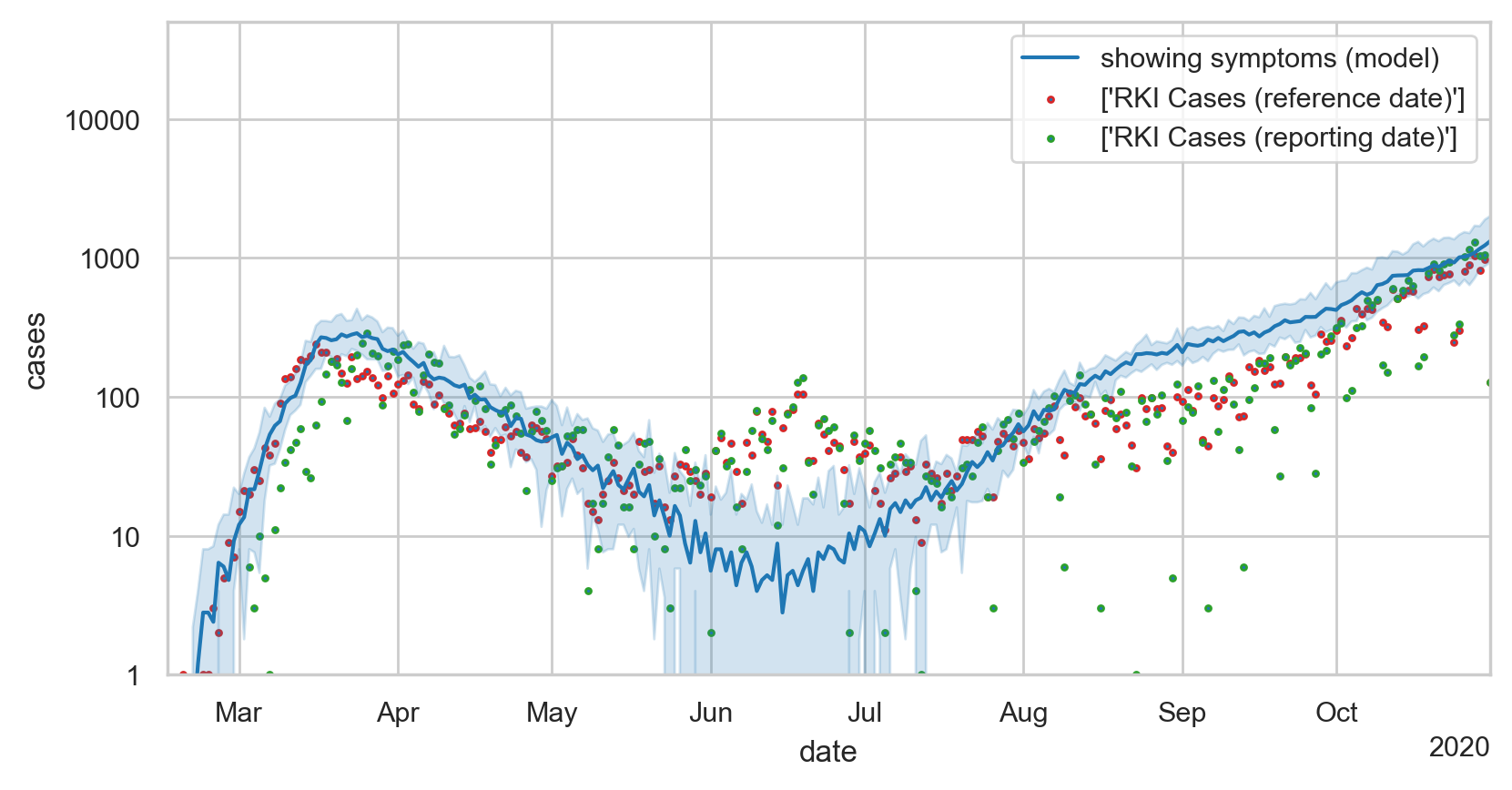}}
\centerline{\includegraphics[width=0.8\textwidth]{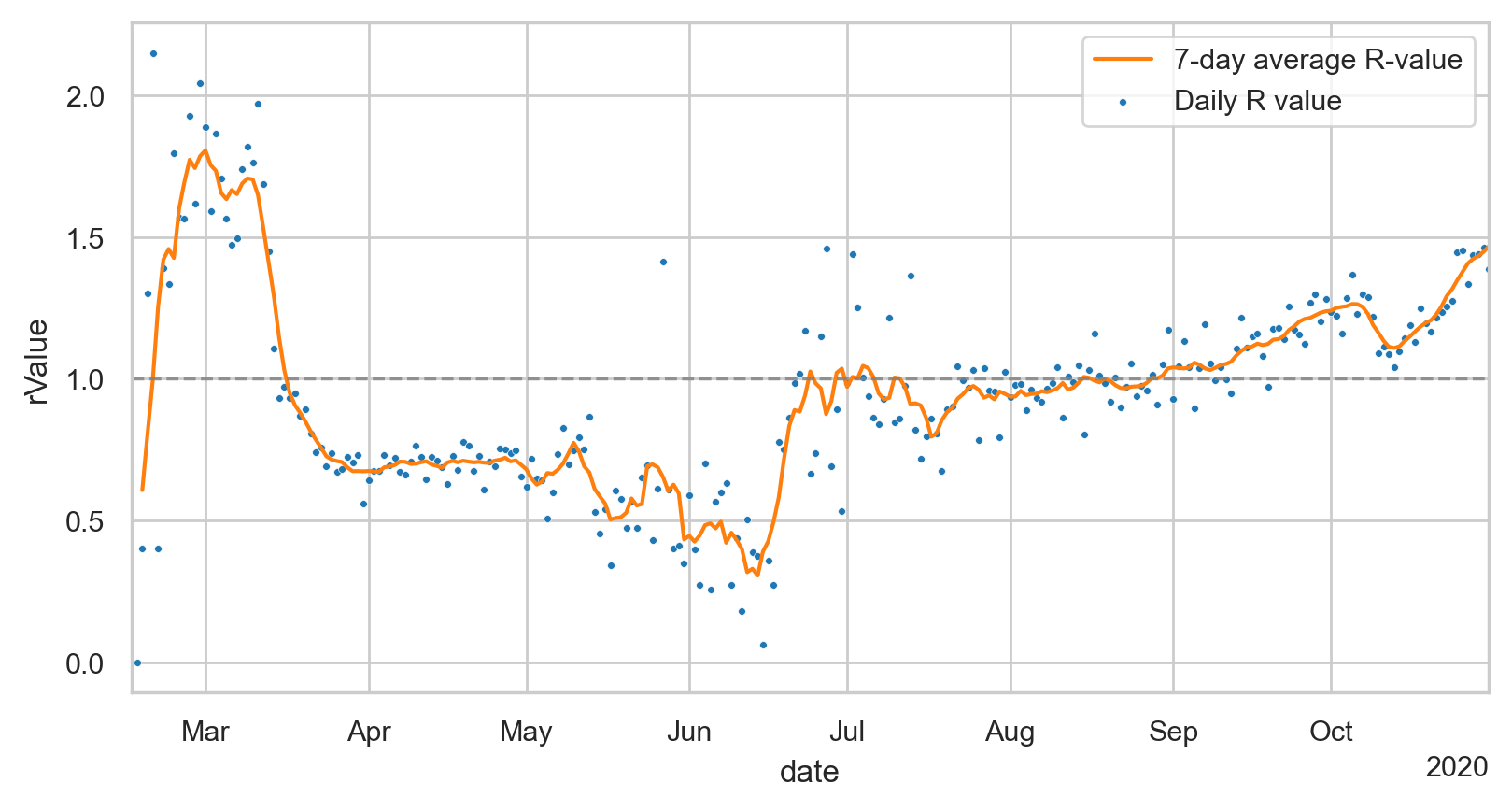}}
\caption{TOP: RKI case numbers (red: reference date $=$ reported day of symptoms onset; green: reporting date $=$ date when local health department was notified; \cite{Robert_Koch-Institut2020-tx}) together with final calibration result (averaged over 10 runs). 
BOTTOM: Resulting reinfection rate $R$ (in simulation). Each daily value is averaged over the 10~runs; the orange line in addition averages over seven days.
The maybe unexpected drop in October stems from the school vacations (see Fig.~\ref{fig:infections-per-activity}).  --  For more information see \url{https://covid-sim.info/2020-11-03/sensitivityAggr}.}
\label{fig:infections}
\end{figure}

The simulation is calibrated against the Berlin case numbers (Fig.~\ref{fig:infections}).  COVID-19 is a notifiable disease, and the notifications are collected and published by the Robert Koch Institute (RKI) \cite{Robert_Koch-Institut2020-tx}.  Each record contains at least two dates: The date when the record reaches the local health department
(reporting date), and the date when symptoms
started, called reference date.

In principle, the reference date would be easier to compare with our simulations, since it corresponds to the onset of our \emph{showingSymptoms} state.  Unfortunately, however, it is not clear how reliable that date is.  The health department becomes aware of cases once they are tested positively.  The positive test result becomes available about 2~days after the probe was taken.  The health authorities thus have to connect a positive test with the person, and query the person about when symptoms started.  Self-reported dates of symptoms onset are presumably rather unreliable, in part because of recall errors, in part because what a symptom is is not sharply defined.  The reliability may be improved by using expert interviewers, but those may not always be available.  In addition, when tests are taken from pre- or asymptomatic cases, a date of symptoms onset is not yet available, and for asymptomatic cases never will be.  In such cases, the reporting data is also entered as reference date,
which for pre-symptomatic cases is too early.  Finally, many records are reported completely without this reference data.  RKI provides a procedure to impute the missing reference date \cite{An_der_Heiden2020-ci}, but has to rely on the statistical distribution of the cases where a reference date exists, which may not be a valid assumption since, say, locations that are under stress of high infection numbers may both not enter the reference date \emph{and} receive the test results with additional delay.

In consequence, we decided to plot the case numbers both by reporting and by reference date for comparison.
The result is shown in Fig.~\ref{fig:infections} (top), where the blue line traces the number of new cases with state \emph{showingSymptoms} from our simulation.  Fig.~\ref{fig:infections} (bottom) shows the resulting reinfection rate $R$.  To obtain that number, the reinfections caused by each synthetic person in the simulation are registered backwards to the date where that person turned contagious, and then averaged over all persons turning contagious on that date.

\begin{figure}[!h!t!]
\centering
\includegraphics[width=0.8\textwidth]{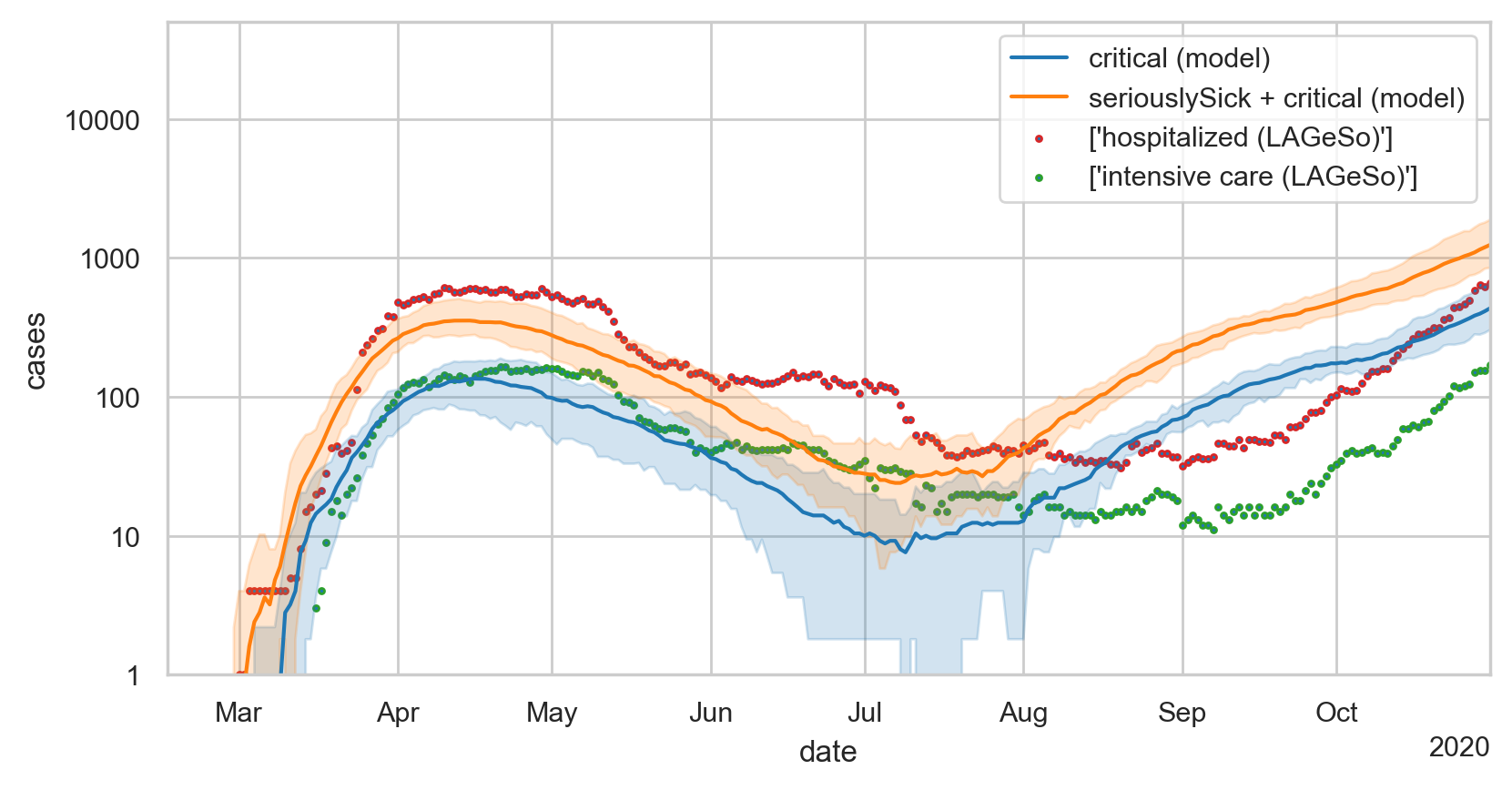}
\caption{Reported hospital cases \cite{Tagesspiegel2020-qw,DIVI_eV_undated-ja} together with simulated results.
}
\label{fig:hospitalNumbers}
\end{figure}

In terms of calibration, the initial growth is, within limits, insensitive against changes of $\Theta$, since it is dominated by the disease import.  This can be explained by the fact that the exponential growth was running ahead in other areas, and in consequence the \emph{share} of infected persons from those areas also grew exponentially.  Only after travel was stopped, disease import also stopped, and the dynamics in Berlin was dominated by internal processes.  
What \emph{is} sensitive against $\Theta$ is the downward slope around April.  In consequence, we adjust $\Theta$ such that the downward slope in the logarithmic plot is reproduced, given the activity reductions and mask compliance provided by our data.  The result is also compared against hospital numbers (Fig.~\ref{fig:hospitalNumbers}), which confirms our calibration. 

The case numbers over the summer contain one large outbreak in a religious community, where we would claim that out model does not pick up such special cases, or at best as the possibility of large fluctuations in certain regimes.  Otherwise, over the summer, the reinfection rate $R$ was below one (Fig.~\ref{fig:infections} bottom).  In the middle of July, the infection dynamics starts picking up again.  The corresponding $R$, however, is mostly below one.  Presumably, this is the consequence of the disease import (cf.\ Fig.~\ref{fig:disease-import}) -- note that our $R$ is taken directly from the simulation, i.e.\ we register backwards at the date when a person turns contagious 
how many persons it will infect later.  That is, there are increasing case numbers because there is disease import, but it does not go along with an $R$ that is larger than one.

\label{sec:comp-with-hosp}

\subsection{Infections per activity type}

\begin{figure}[!h!t!]
\centering
\includegraphics[width=0.8\textwidth]{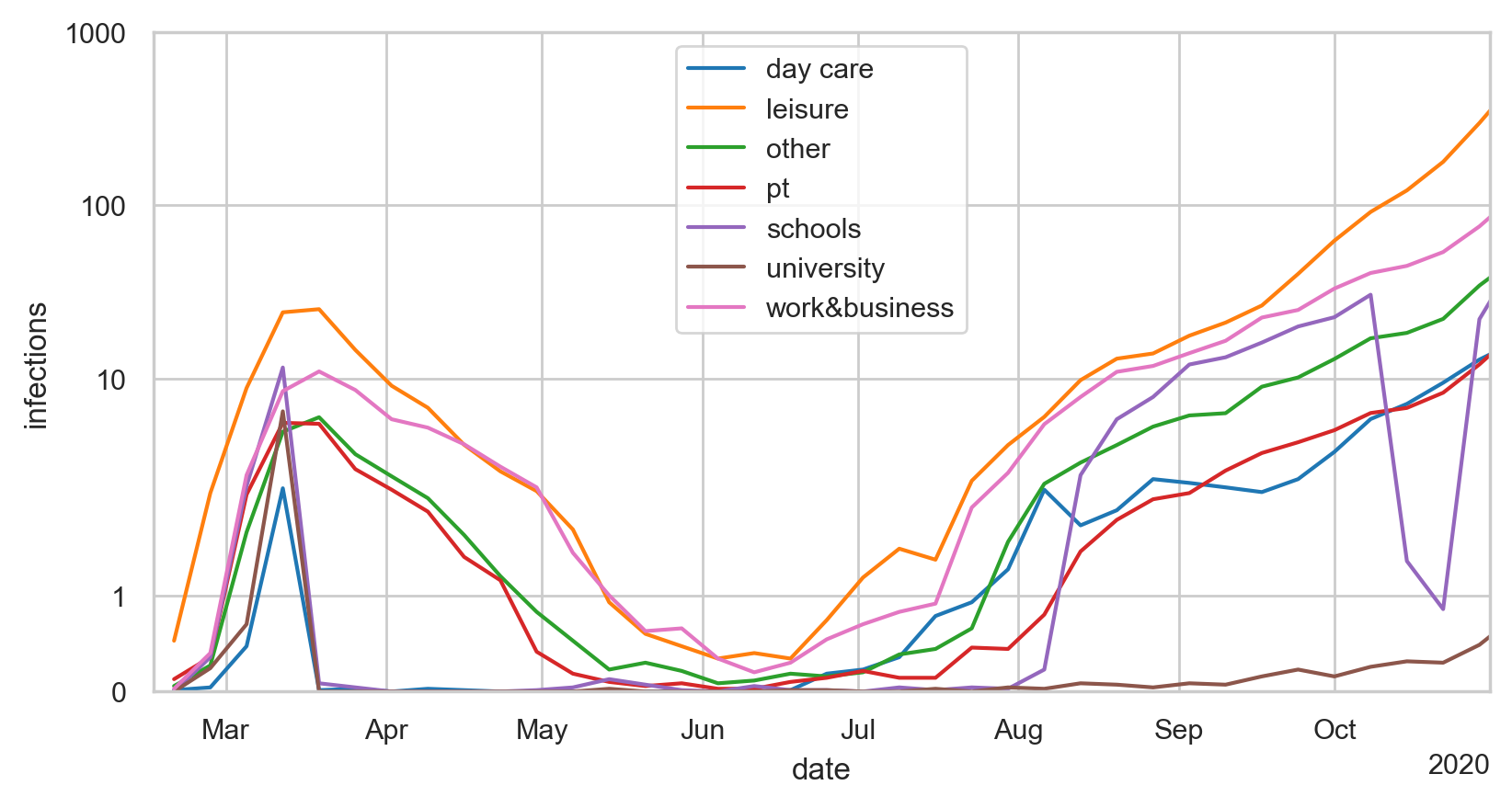}
\includegraphics[width=0.8\textwidth]{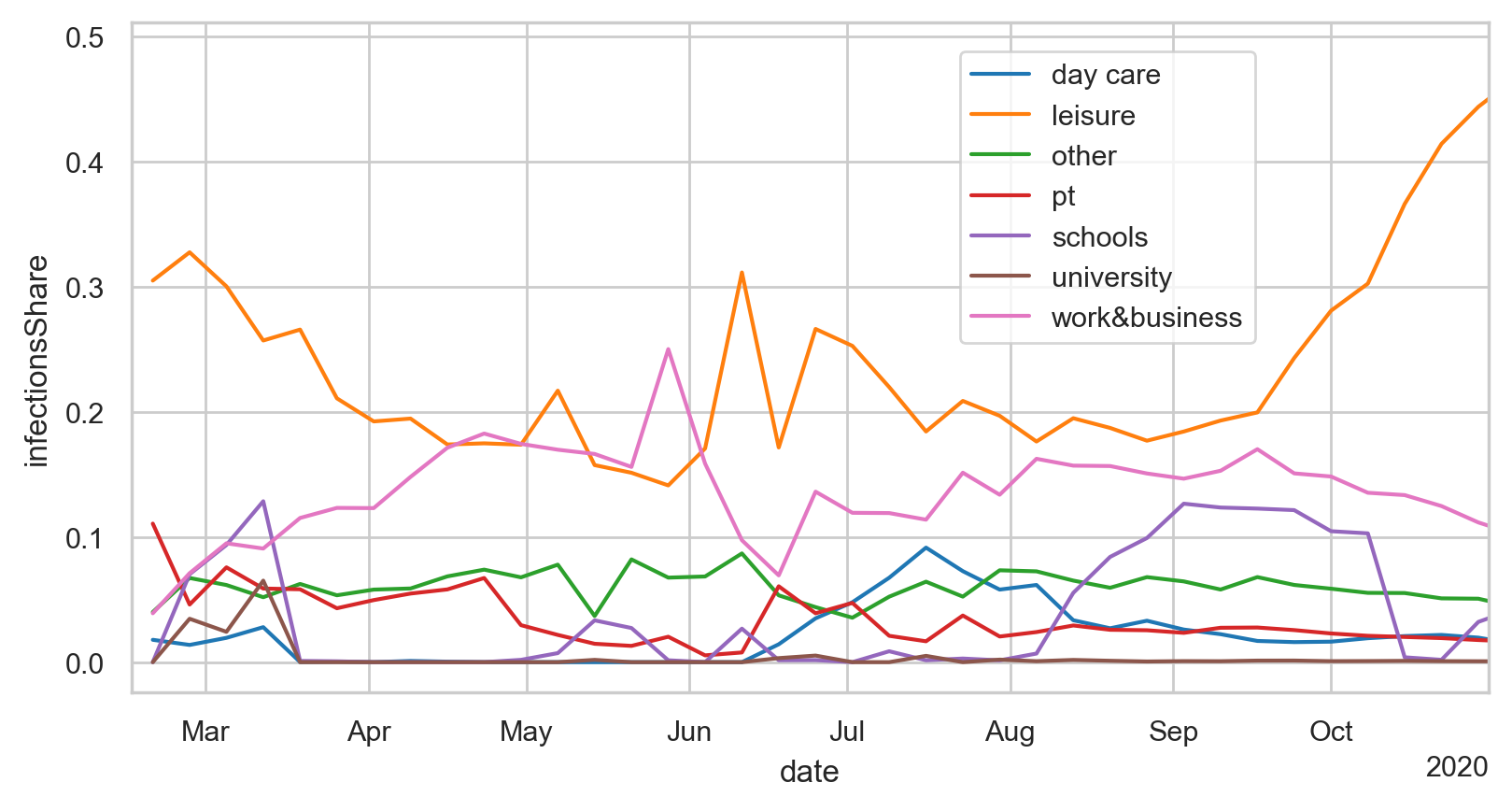}
\caption{TOP: Infections per activity type.  Note logarithmic scale.  BOTTOM: Share of infections per activity type. 
The values are averaged over the same 10~runs as for the other figures, and in addition aggregated into weekly bins.  One can see, for example, the return to school near the beginning of August, and the fall vacations in October.  The shares ($=$ bottom figure) are highly unstable around June because of low infection numbers.}
\label{fig:infections-per-activity}
\end{figure}

Evidently, in our microscopic models we can track how many infections happen at which activity type.  Fig.~\ref{fig:infections-per-activity} shows, on top, the absolute numbers of infection types for the simulation, 
and below the \emph{share} of infections per activity type over time. Initially, all activity types play a role.  
After the closure of the universities, schools, and day care in March, both their absolute numbers and their shares went to zero.  
At the same time, the infections share of work (pink) in April and May reflects that persons were drifting back to normal activity patterns (cf.\ Fig.~\ref{fig:reducedParticipation2}).  
Leisure (orange) would have shown the same trend, but that was counter-acted by the increasing shift of activities to outdoors.   
In the bottom plot, the red line shows how the share of infections in public transit decreases significantly near the end of April because of increased wearing of masks.  (Recall that we use observed mask compliance.)
The \emph{shares} in June cannot be interpreted because the absolute numbers are too low.  In July we see how day care (blue) picks up, because it was re-opened.  Schools re-open in the second week of August, and pick up accordingly (purple).  Also, two weeks of school vacation in October are clearly reflected in the purple curve.
From September on we then see a strong increase of the infections share of leisure activities -- corresponding to moving leisure activities from outdoors to indoors as explained in Sec.~\ref{sec:outdoors-vs-indoors}.

The results point to the importance of reducing infections during leisure activities if one wants to keep the dynamics under control during winter.  After that, school and work have roughly equal weight.  The following interventions would keep these under control:
\begin{itemize}
\item For work, it is suggested to either only work in single-person offices, or wear (N95) masks also at the workplace.  According to our computations, this would reduce the contribution of work activities to the overall dynamics to no longer relevant. 
\item For school, it is suggested to combine (N95) masks with dividing classes by two and having them attend school only on alternating days.  According to our computations, this would reduce the contribution of school activities to the overall dynamics to no longer relevant.
\end{itemize}

Germany has introduced restrictions for leisure activities starting Nov/2.  In our mobility data, we accordingly see reductions, but they will at best reduce the share of the leisure infections by a factor of two.  This means that it will remain the activity type with the largest share.  The interventions for work and school have been recommended, but have not been implemented widely.  Our current expectation \cite{Muller2020-wa} is that these measures in the leisure sector will be sufficient to stop the (exponential) growth, but they will not be enough to allow for a quick decline of the infection numbers.  Additional measures will be needed to achieve that.

\paragraph{Intuition for these results}

In an older version of the model \cite{Muller2020-qm}, we had all contact intensities set to one. The contributions of each activity type to the infection dynamics then in first order corresponded to the average weekly time consumption in the respective activity.  For example, averaged over the week, school consumes about 5~hours per day for persons going to school.  However, since in Berlin only about 10\% of the population are school children,\footnote{%
\url{https://www.statistik-berlin-brandenburg.de/BasisZeitreiheGrafik/Bas-Schulen.asp?Ptyp=300&Sageb=21001&creg=BBB&anzwer=5}
} the average time consumption for the school activity is only 0.5~hours per day when taken across the whole population.  In contrast, there are more persons going to work than to school, thus increasing the weight of work in the infection dynamics.  The by far largest weight, however, comes from the leisure activities, which are not necessarily more hours per week for each individual person, but where \emph{all} persons contribute to this type of time consumption.  In consequence, restricting leisure activities has a large effect.

In the present model, this is now multiplied with the normalized contact intensities, cf.~Tab.~\ref{tab:contactIntensities}.   In consequence, leisure, which already had a large share before, now gets even more weight.  Work, despite occupying similar amounts of time, is weighted down because of the much smaller normalized contact intensity. On the other end of the scale, public transport has a high normalized contact intensity, but the times spent in public transport are considerably smaller.

A complicated case are schools and day care: They occupy large amounts of time, \emph{and} have a large normalized contact intensity, both somewhat similar to leisure.  In consequence, the re-opening of day care in July and of the schools in August should have had strong consequences in the infection numbers (Sec.~\ref{sec:reduced-infectivity-etc} in the appendix, in particular Fig.~\ref{fig:noAgeDepInfModel}).  We took the observation that that did not happen as confirmation that their larger-than-average contact intensity is compensated for by a smaller-than-average infectivity and susceptibility (cf.\ Sec.~\ref{sec:children}).  Clearly, this is specific to (our current understanding of) COVID-19.
For other diseases, for example influenza, children may have a larger infectivity/susceptibility than adults, which then multiplied with their large contact intensity would lead to a large contribution to the infection dynamics.  In consequence, these sub-models need to be understood and re-calibrated for each individual communicable disease.

\section{Discussion}
\label{sec:discussion}

\paragraph{Comparison to compartmental models}

Arguably, compartmental models are the mainstay of epidemiological modelling.  Our approach, in contrast, follows individual synthetic persons.  
These individual persons can be enriched by person-centric attributes such as age or individual risk factors.
Disease progression is individual, taking into account these demographic and other person-centric attributes.
Similar to compartmental models, the base reinfection rate and the starting date need to be calibrated from case numbers.  
However, both the spatial and the social interactions in our model come directly from data.  
Also, behavioral reductions in activity participation come directly from data.  
Mechanical aspects such as the wearing of masks by certain persons and/or at certain activity types can be integrated very simply into the model, by reducing virus shedding, virus intake, or both.  
Travel in public transport is already integrated.
Organizational suppression approaches, such as contact tracing, can be simulated mechanically, thus extracting information about the allowed delays between symptom onset and reaching contacts, the failure rate, etc.

We were able to bring this up quickly: Coding of the infection code was started at the end of Feb/2020; our first preprint is from 20/Mar/2020 \cite{Muller2020-wv}; our first report to the government is from 8/Apr/2020 \cite{Muller2020-vq}; we have reported to the government regularly since then\footnote{Cf.~\url{https://depositonce.tu-berlin.de/simple-search?query=modus-covid}}.  Evidently, we were drawing from our experience and expertise with person-centric travel models.  Still, it means that given the right experience and data availability, the method is not overly heavyweight, and then has many advantages over compartmental models.

The basic behavior of the model is like that of any S(E)IR model, i.e.\ exponential growth until a sufficient share of the population is immune, followed by exponential decline (cf.~blue line in Fig.~\ref{fig:infections}).  Also the beginning and the speed of the growth are calibrated in similar ways.
In most models, however, interventions such as reductions in out-of-home activity participation, masks, or contact tracing, need to be parametrized into parameter changes of the S(E)IR model, most notably the infection rate \cite{Dehning2020-th,Eichner_undated-ao,Neher_undated-ts,Althaus2020-jk}.  The only models that use human activity patterns directly that we are aware of are the three models described in \cite{Halloran2008-ag}.  Out of these, we are aware of an application to COVID only by the Imperial College model \cite{Ferguson2020-lk}.  Their results are roughly in line with ours.  That model, at the time, used a doubling of cases every 10~days; reality, with a doubling every 3 days, was possibly even more dramatic than their predictions.  However, their model was purely predictive, i.e.\ other than us they did not use mobility data to gauge the actual reductions in activity participation.

\paragraph{Under-reporting}
\label{sec:underreporting}

A known issue with epidemiological data and thus the simulations that build on it is the issue of under-reporting, i.e.\ that there are more cases in reality than are in the data.  For our model, this would imply to ``raise'' the curve of infections, e.g.\ in Fig.~\ref{fig:infections}, to a higher level.  In order to achieve this, we could, for example, feed the model with initial seeds at some earlier point in time.  This would, however, lead to a lower slope in the log-plot for the early days, in contrast to the data, and thus does not seem plausible.  An alternative would be to assume that the disease import which drives our initial phase, Fig.~\ref{fig:disease-import}, is itself under-reported.  This would be entirely plausible.  This would also increase the hospital numbers, Fig.~\ref{fig:hospitalNumbers}, which would make the curve of \seriouslySick\ more realistic, but make the curve of \critical\ less realistic.  Also, other studies point to relatively little under-reporting in Germany \cite{Fraunhofer-Institut_fur_Techno-_und_Wirtschaftsmathematik_ITWM2020-so,Robert_Koch_Institute2020-gf}.  As long as the number of sero-positive persons in Germany remains in the single-digit percentage ranges \cite{Robert_Koch_Institute2020-gf}, the simulation is not strongly affected by this issue.

\paragraph{Predictions}
\label{sec:predictions}

The model is used for predictions.  We decided to not add them into the paper since any prediction we make now would be historical quickly.  Our regular reports to the government, and thus our predictions, all have a DOI, for example \cite{Muller2020-vq} or \cite{Muller2020-wa}.\footnote{%
Again, see \url{https://depositonce.tu-berlin.de/simple-search?query=modus-covid}.
}

\section{Conclusions}
\label{sec:conclusions}

We combine a person-centric human mobility model with a mechanical model of infection and a person-centric disease progression model into an epidemiological simulation model.  Different from other models, we take the movements of the persons, including the intervening activities where they can interact with other people, directly from data.  For privacy reasons, we rely on a process that takes the original mobile phone data, extracts statistical properties, and then synthesizes movement trajectories from the statistical properties;  one could use the original mobile phone trajectories directly if they were available.  
The model is used to replay the epidemics in Berlin.  This allows important insights into the societal transmission from government actions to mobility behavior to infection dynamics.  Importantly, it turns out that the population started reducing its out-of-home activities \textit{before} the government asked/ordered the population to do so.  The model is then used to evaluate different intervention strategies, such as closing educational facilities, reducing other out-of-home activities, wearing masks, or contact tracing, and to determine differentiated percentage changes of the reinfection number $R$ per intervention. 

\section*{Availability of data and materials}

 For computer code see \url{https://github.com/matsim-org/matsim-episim}. Simulations were computed with version 0683bd27d80963e16af95395790959ae5e1578a0 of the code, started with command 
 \begin{verbatim}
 java -jar matsim-episim-1.0-SNAPSHOT.jar runParallel \ 
      --setup org.matsim.run.batch.BerlinSensitivityRuns \ 
      --params org.matsim.run.batch.BerlinSensitivityRuns$Params
\end{verbatim}
Some of the input data, in particular the synthetic mobility traces, are currently under a license, but we are in the final stages of negotiations with the data provider and are confident that they can be made available soon.
 
The output data used for the figures can be retrieved at \url{https://svn.vsp.tu-berlin.de/repos/public-svn/matsim/scenarios/countries/de/episim/battery/2020-11-03/sensitivity/}.

\section*{Acknowledgements}

We thank Kai Martins-Turner and Dominik Ziemke for discussions.
We are grateful to BVG (Berlin public transit operator) for providing the mask compliance rates which they surveyed on a daily basis.
The work on the paper was funded by the Ministry of research and education (BMBF) Germany (01KX2022A) and TU Berlin; regular reports can be found through this search: \url{https://depositonce.tu-berlin.de/simple-search?query=modus-covid}. Zuse Institute Berlin (ZIB) provided CPU time.

\bibliographystyle{vancouver}%
\bibliography{paperpile,tub}     %

\appendix
\section{Appendix}
\subsection{Sensitivities}
\label{sec:sensitivities}

This section discusses what happens when certain elements of the model are switched off.

\subsubsection{Disease import}

\begin{figure}[!h!t!]
\centerline{%
\includegraphics[width=0.7\hsize,trim=0 0 0 0,clip]{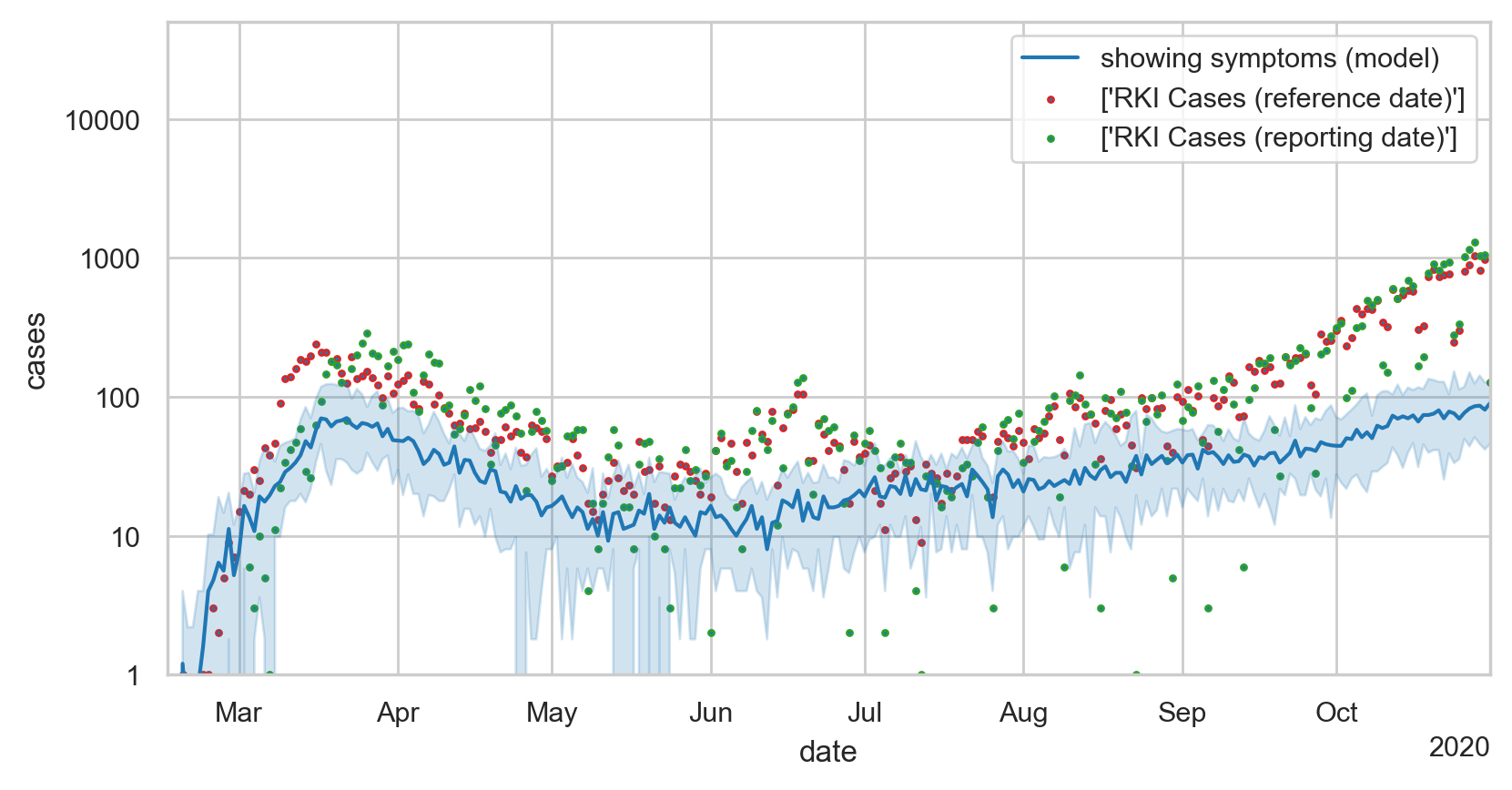}
}
\centerline{%
\includegraphics[width=0.7\hsize,trim=0 0 0 0,clip]{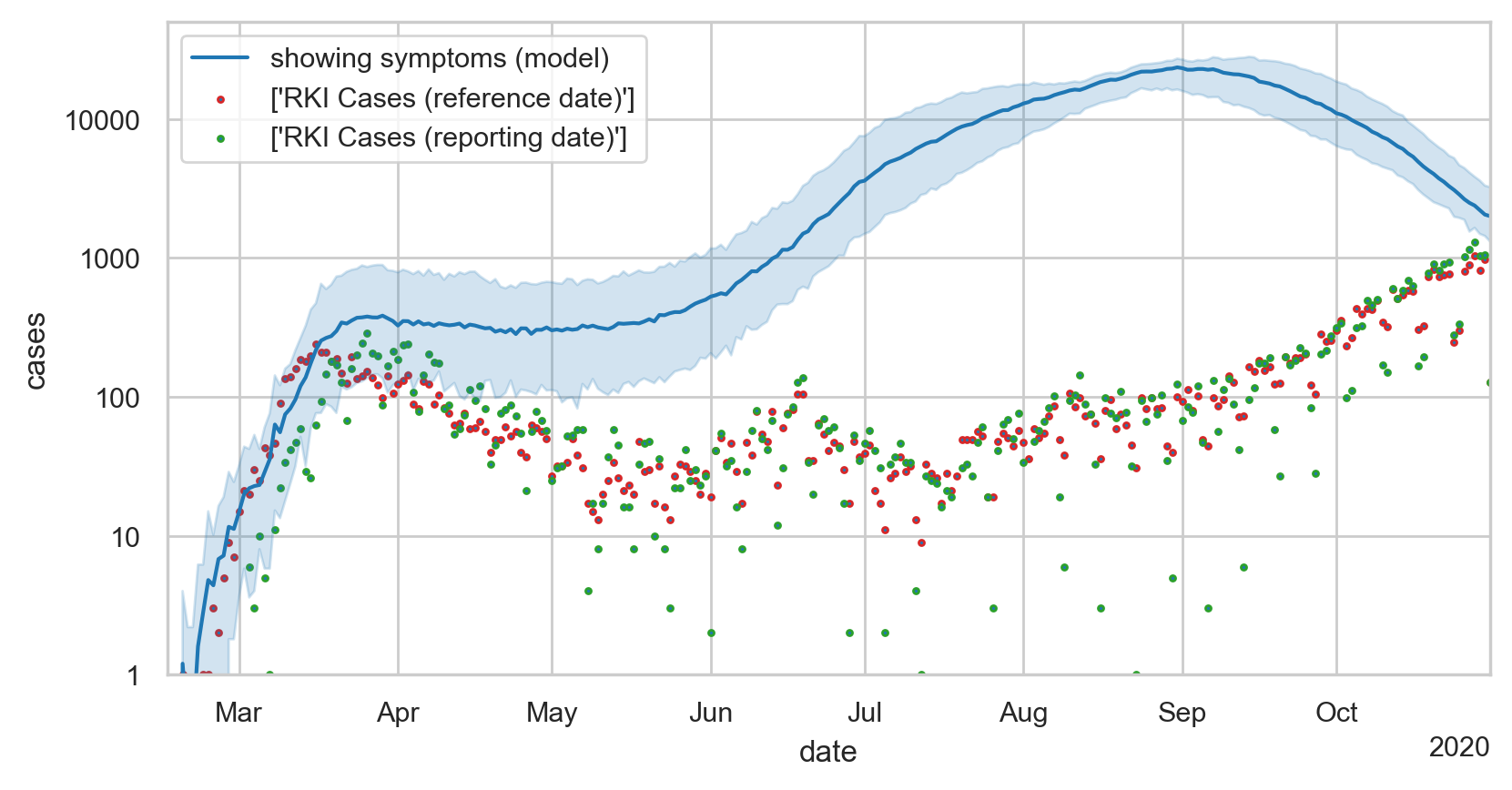}
}
  \caption{Model where the time-dependent disease import (Fig.~\ref{fig:disease-import}) is replaced with a constant disease import (4 persons per day = 1 person per day in 25\% sample).  
TOP: The infection numbers come out too low. BOTTOM: 
After re-calibrating $\calibrationParam$ to the initial growth in March, the infection numbers afterwards come out too high.}
\label{fig:noDiseaseImport}  
\end{figure}

Fig.~\ref{fig:noDiseaseImport} (top) shows what would happen if disease import was not included into the model.  Since now the initial growth of the epidemics is too slow, the calibration parameter $\calibrationParam$ is re-calibrated so that the initial growth is reproduced correctly (Fig.~\ref{fig:noDiseaseImport} bottom).  One clearly notices that this results in much more infection activity in the model than in reality.  That is, according to our model the fast growth in March in Berlin was partially caused by disease import; without disease import, we need to use a $\Theta$ and thus a re-infection rate that is too high for what follows afterwards.

\subsubsection{Masks}

\begin{figure}[!h!t!]
  \centerline{%
\includegraphics[width=0.7\hsize,trim=0 0 0 0,clip]{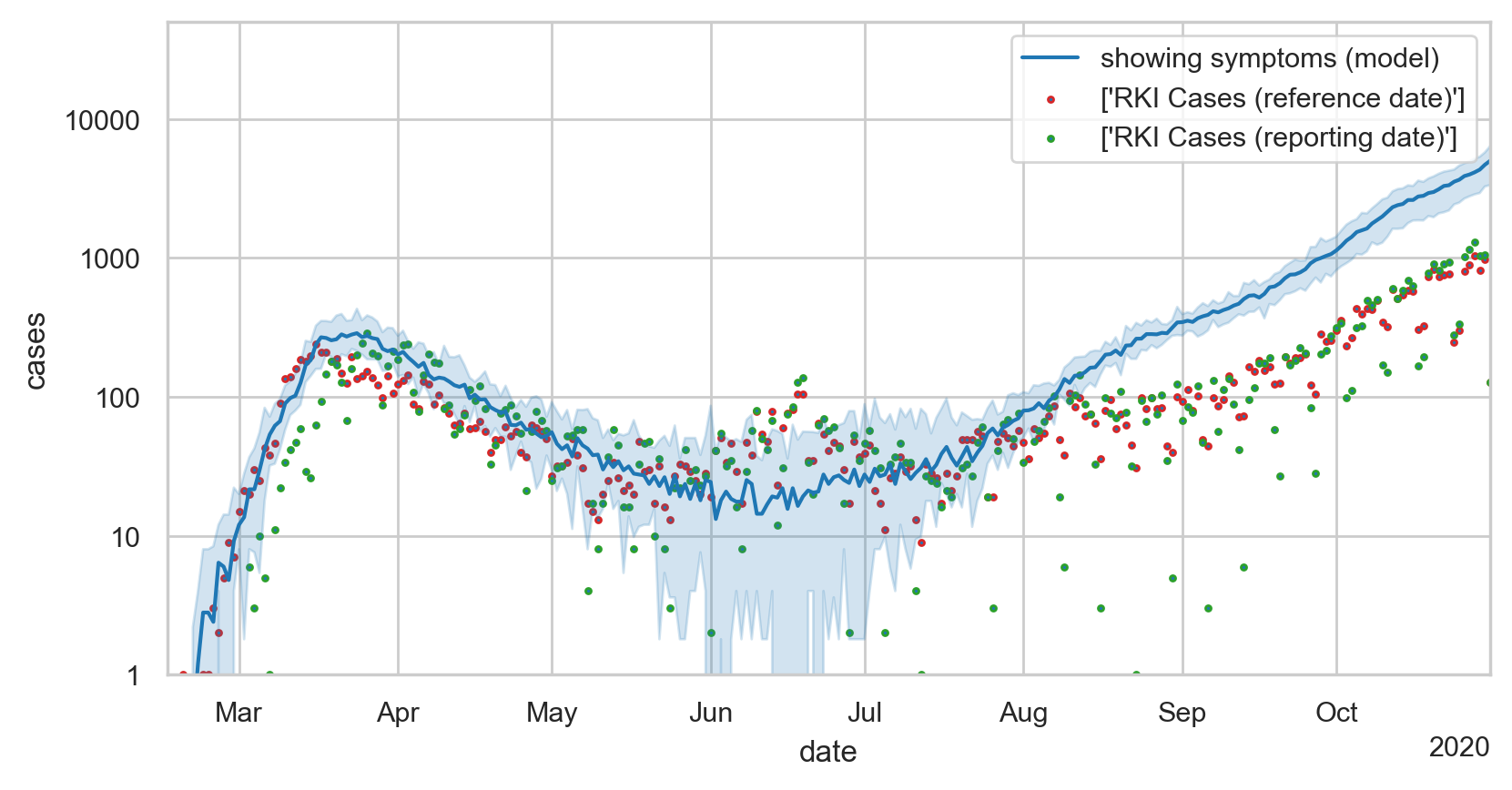}
}
  \caption{Model without masks during shopping/public transport.  The calibration parameter $\calibrationParam$ is \emph{not} recalibrated.}
  \label{fig:noMasks}  
\end{figure}

Fig.~\ref{fig:noMasks} shows the consequence of wearing masks in public transport and during shopping.  There is a noticeable difference to Fig.~\ref{fig:infections}, but the difference is not huge.  

\subsubsection{Contact tracing}

\begin{figure}[!h!t]
  \centerline{%
\includegraphics[width=0.7\hsize,trim=0 0 0 0,clip]{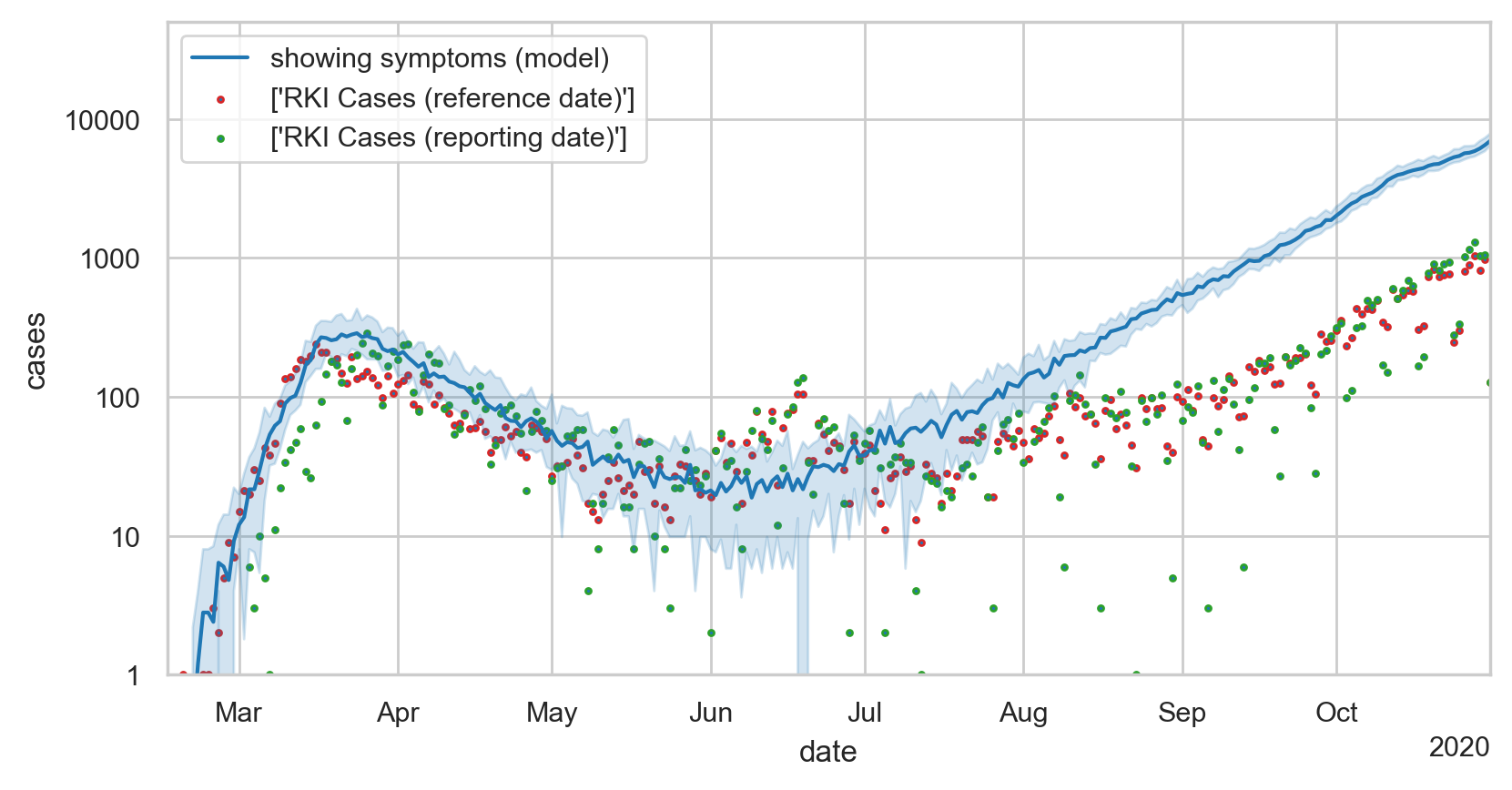}
}
  \caption{Model without contact tracing.  The calibration parameter $\calibrationParam$ is \emph{not} recalibrated.}
\label{fig:noTracing}
\end{figure}

Fig.~\ref{fig:noTracing} shows the consequence of not doing contact tracing.  In consequence, the infection numbers start growing already in June, and at the beginning of November would be considerably larger than they actually are.  One also notices that there is now a uniform slope in the logarithmic plot from July to November, while with contact tracing (Fig.~\ref{fig:infections}) one clearly sees the increase in slope in October when the contact tracing becomes overwhelmed.

\subsubsection{Reduced infectivity and susceptibility for children}
\label{sec:reduced-infectivity-etc}

\begin{figure}[!h!t]
  \centerline{%
\includegraphics[width=0.8\hsize,trim=0 0 0 0,clip]{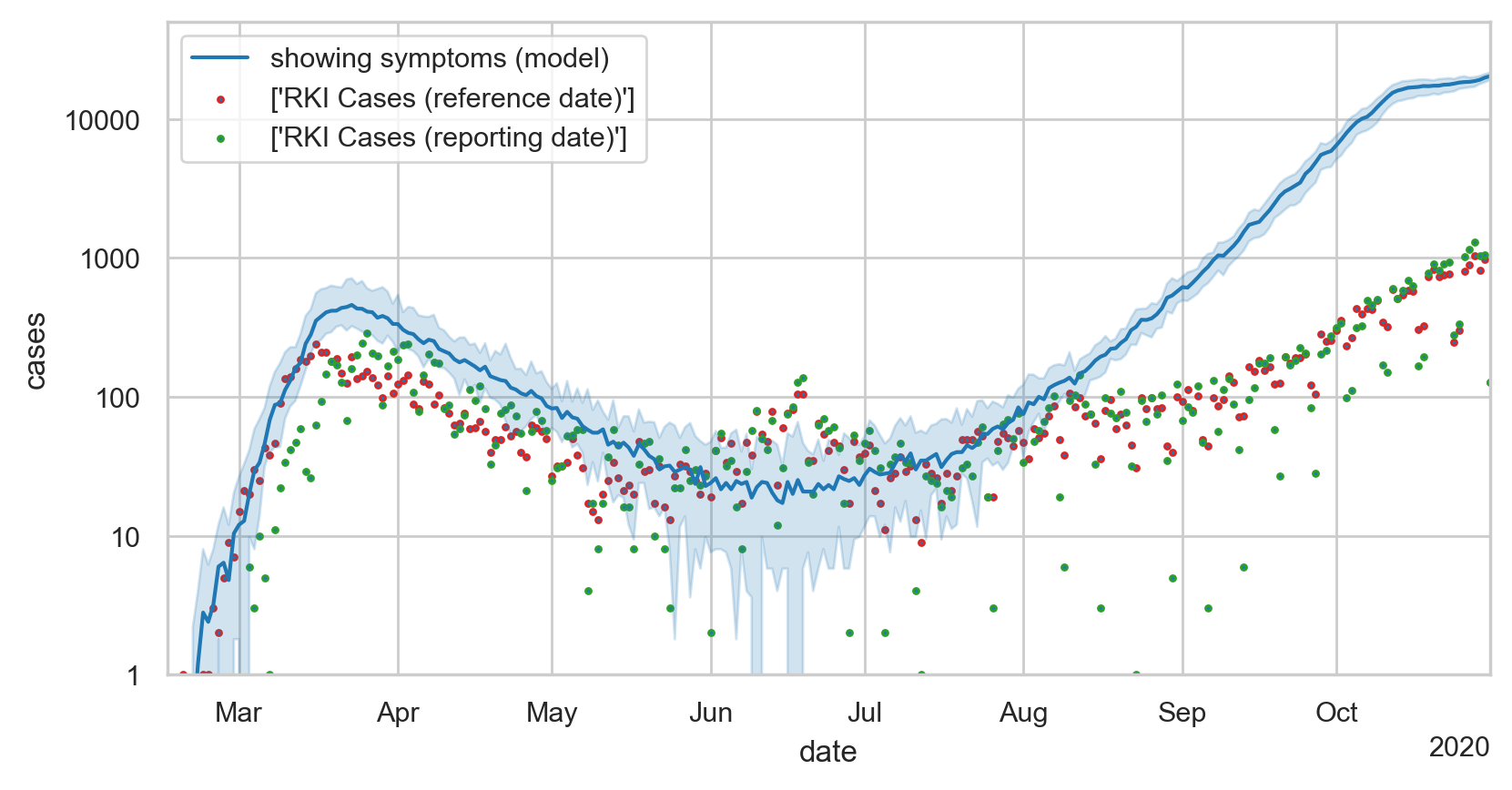}
}
  \caption{Model where infectivity and susceptibility of children are not reduced.  The calibration parameter $\calibrationParam$ is \emph{not} recalibrated.}
  \label{fig:noAgeDepInfModel}
\end{figure}

Fig.~\ref{fig:noAgeDepInfModel} shows the simulation without reducing the infectivity and susceptibility for children (Sec.~\ref{sec:children}).  Schools have re-opened on Aug/8; not applying these two reductions results in an infection dynamics that is much stronger than what was observed in reality.

\subsubsection{Closing of the educational sector}

\begin{figure}[!h!t!]
  \centerline{%
\includegraphics[width=0.8\hsize,trim=0 0 0 0,clip]{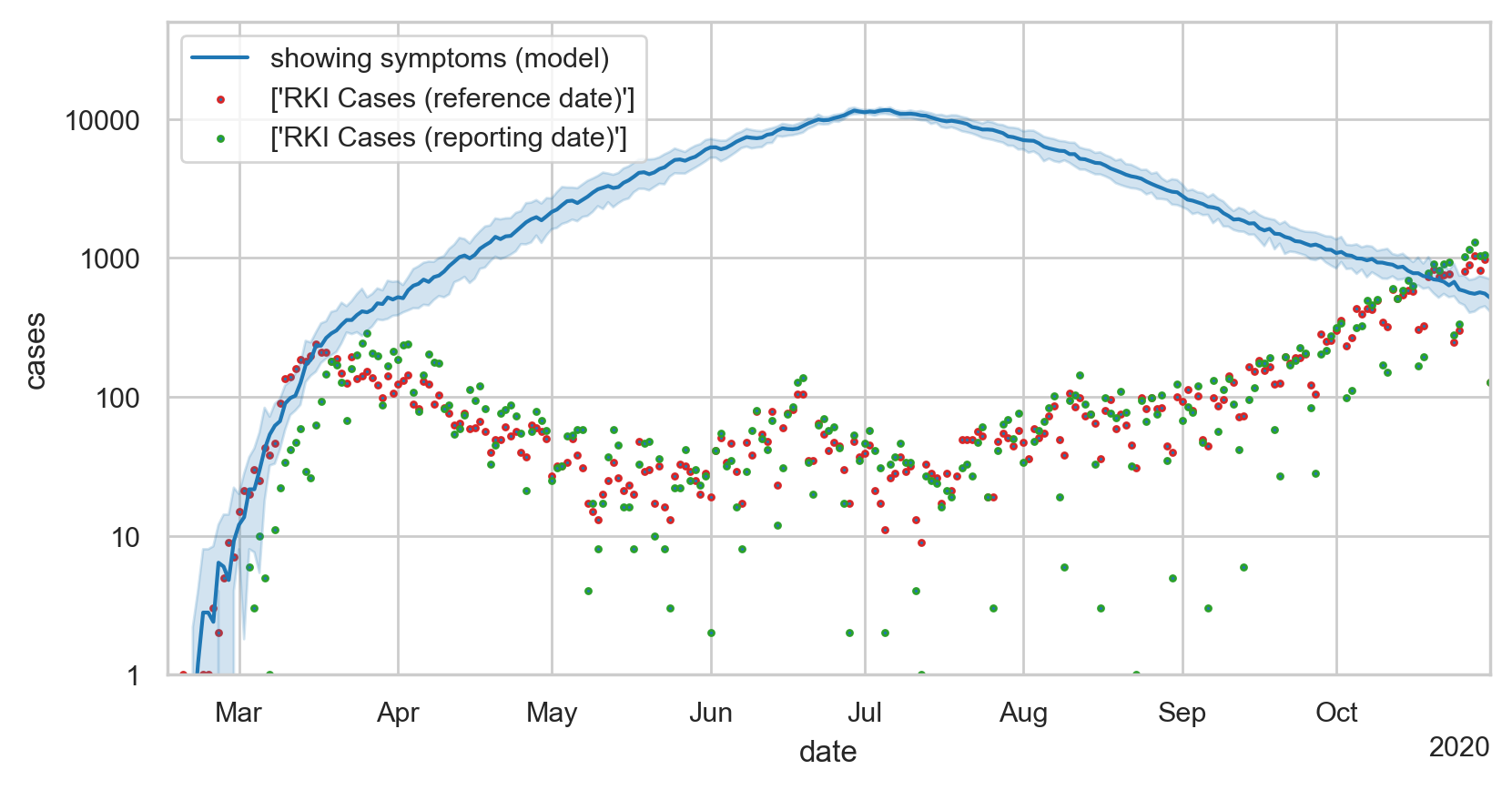}
}
  \centerline{%
\includegraphics[width=0.8\hsize,trim=0 0 0 0,clip]{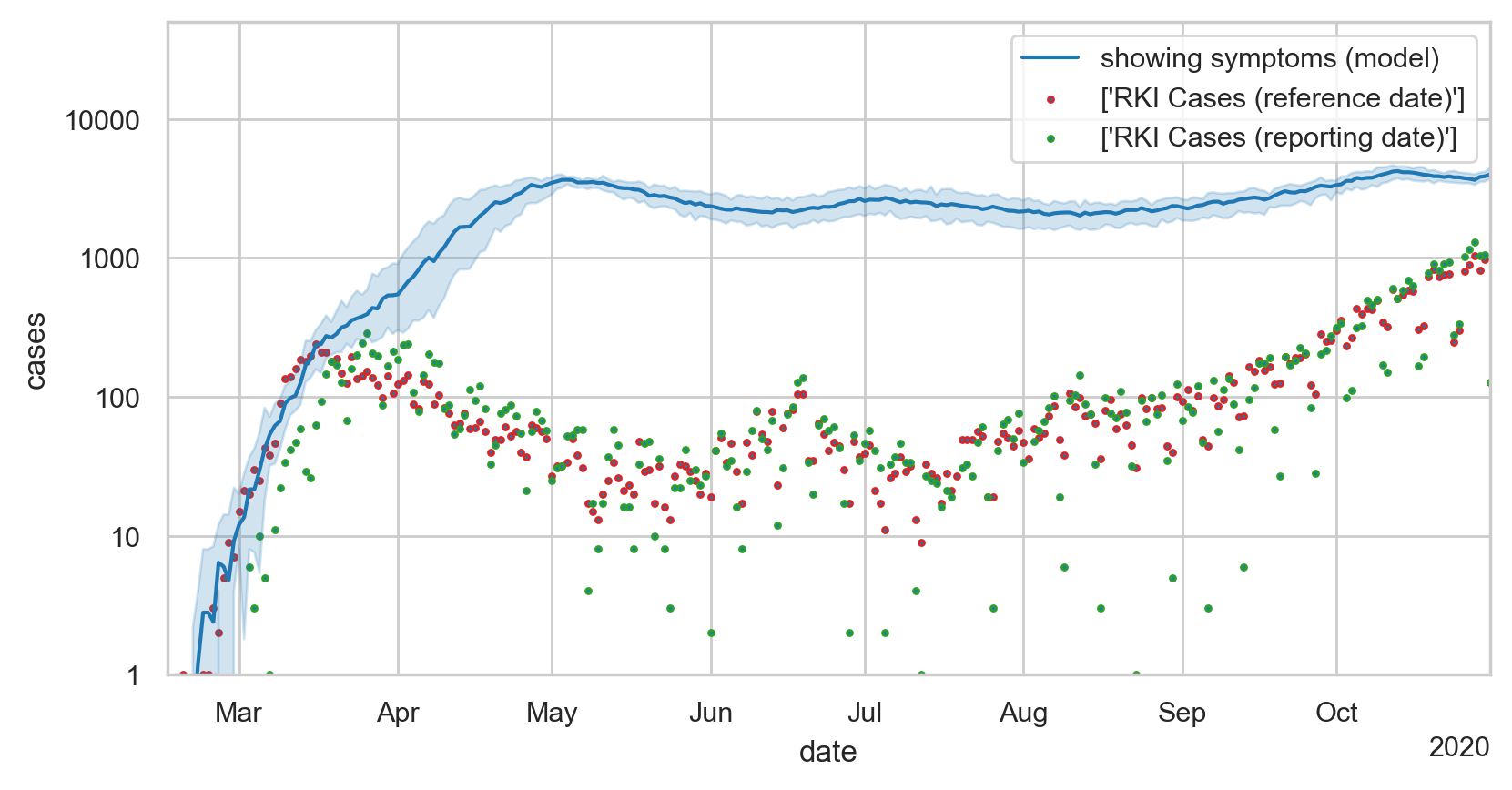}
}
\centerline{%
\includegraphics[width=0.8\hsize,trim=0 0 0 0,clip]{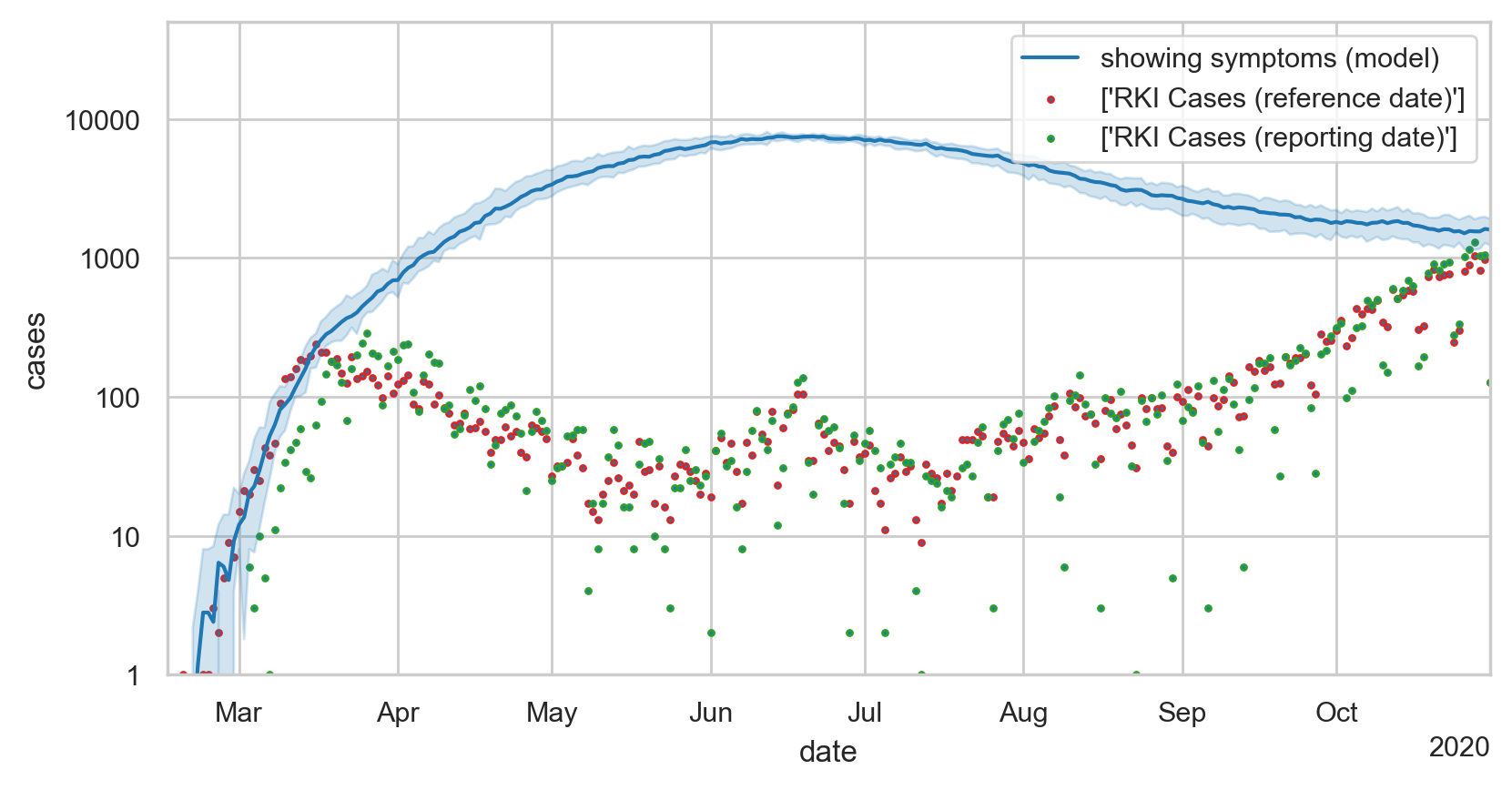}
}
\caption{TOP: Model without restrictions in schools and day care.  MIDDLE: Model without restrictions in universities.  BOTTOM: Model without restrictions outside of the educational sector.}
\label{fig:educationalSector}
\end{figure}

Figs.~\ref{fig:educationalSector} (top) shows what would have happened if schools and day care had been kept fully open during the initial restrictions in March;
Fig.~\ref{fig:educationalSector} (middle) shows what would have happened if universities had been kept fully open during the initial restrictions in March;
Fig.~\ref{fig:educationalSector} (bottom) shows what would have happend if \emph{only} the educational sector had been closed in March.  One finds that neither measure by itself would have been sufficient to bring infection numbers down during the first wave; only the combination of closing day care, schools, universities and strongly reducing all other activities was strong enough to bring infection numbers down.

\subsubsection{Discussion}

We conclude that, given current knowledge, all the elements of our model are necessary to track the development so far.

\clearpage

\end{document}